\documentclass[amsmath,eqsecnum,preprint,pre,aps,showpacs]{revtex4}
\usepackage{amsmath}
\usepackage{graphicx}

\begin{document}

\title{Generalized Onsager-Machlup's theory of
thermal fluctuations for non-equilibrium systems}
\author{ M. Medina-Noyola}

\address{Instituto de F\'{\i}sica {\sl ``Manuel Sandoval
Vallarta"}, Universidad Aut\'{o}noma de San Luis Potos\'{\i},
\'{A}lvaro Obreg\'{o}n 64, 78000 San Luis Potos\'{\i}, SLP,
M\'{e}xico}
\date{\today}

\begin{abstract}
In this work a generalization of Onsager-Machlup's theory of
time-dependent thermal fluctuations of equilibrium systems is
proposed, to the case in which the system relaxes irreversibly along
a non-equilibrium trajectory that can be approximated as a sequence
of stationary states. This generalization is summarized by a
canonical description of the dependence of the two-time correlation
function $C(t+\tau,t)$, and of the equal-time correlation function
$\sigma(t)\equiv C(t,t)$ (the covariance of the fluctuations), on
the non-equilibrium relaxation time $t$
\bigskip
\bigskip
\bigskip
\vskip2cm

\noindent \emph{\textbf{This paper is dedicated to the memory of
Joel Keizer}}

\bigskip
\bigskip
\bigskip

\end{abstract}


\maketitle

\section{Introduction}\label{I}

In this paper a generalization is proposed of some aspects of
Onsager-Machlup's theory of equilibrium thermal fluctuations
\cite{onsagermachlup1,onsagermachlup2}, to conditions outside the
so-called linear regime of irreversible thermodynamics
\cite{onsager1,onsager2}, in which this theory is assumed to be
universally valid \cite{keizer,casasvazquez0}. Although the original
motivation of this work was to extend a theory of colloid dynamics
to describe the non-equilibrium slow dynamics in glass-forming
colloidal systems \cite{cipelletti1,martinezvanmegen}, the resulting
generalization actually constitutes a general canonical theory to
describe the dependence of the two-time correlation function
$C(t+\tau,t)$ and of the equal-time correlation function
$\sigma(t)\equiv C(t,t)$ (the covariance of the fluctuations) on the
non-equilibrium relaxation time $t$.

The dynamic properties of colloidal dispersions has been the subject
of sustained interest for many years \cite{1,6,2}. These properties
can be described in terms of the relaxation of the fluctuations
$\delta n({\bf r} ,t)$ of the local concentration $n({\bf r},t)$ of
colloidal particles around its bulk equilibrium value $n$. The
average decay of $\delta n({\bf r},t)$ is described by the two-time
correlation function $F_t(k,\tau)\equiv \left\langle \delta n({\bf
k},t+\tau)\delta n(-{\bf k},t)\right\rangle $ of the Fourier
transform $\delta n({\bf k},t)$ of the fluctuations $\delta n({\bf
r} ,t)$, whose equal-time limit is $S_t(k)\equiv F_t(k,\tau=0) =
\left\langle \delta n({\bf k},t)\delta n(-{\bf k},t)\right\rangle $.
One may refer to the time $\tau$ as the \emph{correlation time}, and
to the time $t$ as the \emph{relaxation time}. If some external (or
internal) constraints that kept a system at a certain macroscopic
state are broken at time $t=0$, the system relaxes spontaneously to
its new thermodynamic equilibrium state, and one may also refer to
$t$ as the \emph{waiting} or \emph{ageing} time
\cite{cipelletti1,martinezvanmegen}. If the system has fully relaxed
to, and/or remains in, a thermodynamic equilibrium state, these
properties no longer depend on $t$, i.e., $F_t(k,\tau)=F(k,\tau)$
and $S_t(k)=S(k)$. The equilibrium stationary correlation function
$F(k,\tau)$ is referred to as the intermediate scattering function,
and its initial value $S(k)$ as the static structure factor. These
properties can be measured by a variety of experimental techniques,
including (static and/or dynamic) light scattering
\cite{cipelletti1,martinezvanmegen,1}.

The static structure factor $S(k)$, being a thermodynamic property,
is amenable to theoretical calculation using statistical
thermodynamic methods \cite{mcquarrie}. The fundamental
understanding of $F(k,\tau)$, on the other hand, requires the
development of theoretical methods to describe the diffusive
relaxation of the local concentration fluctuations, and a number of
such approaches have been proposed for their theoretical calculation
\cite{1,6,2,13,cohen,14,15,16}. One of the earliest was developed by
Hess and Klein \cite{6,13}, who translated to colloids the
mode-coupling self-consistent theory of molecular liquids
\cite{goetze3,goetze4}. More recently, N\"{a}gele and collaborators
elaborated further this mode-coupling theory of colloid dynamics
\cite{14,15,16}.

An independent alternative theory of the dynamic properties of
colloidal dispersions has been developed within the last decade, and
is referred to as the self-consistent generalized Langevin equation
(SCGLE) theory of colloid dynamics
\cite{scgle0,scgle1,scgle2,marco1,marco2}. This theory has been
recently applied to the description of dynamic arrest phenomena in
several specific colloidal systems that include mono-disperse
suspensions with hard-sphere interactions, moderately soft-sphere
and electrostatic repulsions, short-ranged attractive interactions,
and model mixtures of neutral and charged particles
\cite{rmf,todos1,todos2,attractive1,soft1,rigo1,rigo2,luis1}.

Until recently, and in spite of the long tradition in the study of
glasses \cite{angell,debenedetti,goetze1}, the only well-established
and successful theoretical framework leading to \emph{quantitative}
predictions of the glass transition was the conventional mode
coupling theory (MCT) of the ideal glass transition
\cite{goetze3,goetze4,goetze1,goetze2}. Many of the predictions of
this theory had been systematically confirmed by their detailed
comparison with experimental measurements in model colloidal systems
\cite{vanmegen1, bartsch, beck, chen1, chen2,pham,
sciortinotartaglia,buzzacaro,sanz1,lu1}. In this context, we can
mention that the more recently-developed SCGLE theory of dynamic
arrest leads to similar dynamic arrest scenarios as MCT \cite{rmf,
todos1} for several specific (mostly mono-disperse) systems,
although for colloidal mixtures important differences may appear in
some circumstances, as reported in Refs. \cite{rigo1,rigo2,luis1}.

An important common feature of both theories in their current status
is that they are able to predict the regions of the control
parameter space in which the system is expected to be a glass, i.e.,
it predicts what we refer to as the ``dynamic arrest phase diagram"
of the system \cite{rigo2,luis1}. Since the predicted state only
depends on the end value of the control parameters, these
predictions only apply to ``reversible" or ``equilibrated" glasses.
Thus, while it is important to pursue the application of these two
theories to specific idealized or experimental model systems and to
compare their predictions, it is also important to attempt their
extension to the description of ``non-equilibrated" glasses, for
which no ``dynamic arrest phase diagram" will make sense without the
specification of the detailed non-equilibrium process leading to the
apparent end state. Ageing effects, for example, should be a
fundamental aspect of the experimental and theoretical
characterization of these non-equilibrium states. These
preoccupations have been addressed in the field of spin glasses,
where a mean-field theory has been developed within the last two
decades \cite{cugliandolo1}. The models involved, however, lack a
geometric structure and hence cannot describe the spatial evolution
of real glass formers. Thus, although computer simulations
\cite{kobbarrat, puertas1} and experimental studies
\cite{cipelletti1,martinezvanmegen} have provided important
information about general properties of ageing, no quantitative
theory is available so far to describe the irreversible formation of
structural glasses.

Almost a decade ago, however, an attempt was made by Latz
\cite{latz} to extend MCT to describe the irreversible relaxation,
including  ageing processes, of a glass forming system after
suddenly driving it into the glassy region of its dynamic arrest
phase diagram. A major aspect of this work involved the
generalization to non-equilibrium conditions of the conventional
equilibrium projection operator approach \cite{berne} to derive the
corresponding memory function equations in which the mode coupling
approximations could be introduced.

In very recent work \cite{eom2} a similar extension of the SCGLE
theory of dynamic arrest has been proposed by the present author. In
contrast with MCT, the fundamental basis of the SCGLE theory does
not involve the use of projection operator techniques. Instead, it
is based on Onsager-Machlup's theory of time-dependent thermal
fluctuations of equilibrium systems. Thus, its extension to
non-equilibrium conditions calls for a generalization to the
nonlinear regime of the general and fundamental laws of linear
irreversible thermodynamics and the corresponding stochastic theory
of fluctuations, as stated by Onsager \cite{onsager1, onsager2} and
by Onsager and Machlup \cite{onsagermachlup1,onsagermachlup2},
respectively, with an adequate extension \cite{delrio,faraday} to
allows the description of relaxation phenomena involving memory
effects. In Ref. \cite{eom2} this generalization of Onsager's theory
was only outlined. The main purpose of the present paper is then to
explain in more detail the main features and the underlying
assumptions of such generalized canonical theory of time-dependent
fluctuations around the irreversibly relaxing state of
non-equilibrium systems.

For this, and since the literature on non-equilibrium extensions of
equilibrium theories is quite diverse and extensive
\cite{keizer,casasvazquez0,casasvazquez1}, and in order to normalize
the basic concepts and scientific context, we emphasize that we
follow to a large extent the philosophical approach of Joel Keizer's
statistical thermodynamic theory of non-equilibrium processes
\cite{keizer}, particularly his account of the Onsager picture
(chapters 1 and 2 of Ref. \cite{keizer}) and some aspects of
Keizer's extension of this picture to non-equilibrium (chapters 3
and 4 of the same reference). However, for our present purpose we do
not find necessary to adhere to Keizer's detailed mechanistic
statistical description in terms of elementary processes.

Thus, taking some elements from Keizer's theory and other elements
not considered by him (such as the non-Markovian extension of
Ornstein-Uhlenbeck processes \cite{delrio}), the present paper
proposes a generalization to nonlinear conditions of some elements
of Onsager's canonical theory. We refer essentially to the
time-evolution equations of the first two moments of the conditional
probability distribution (i.e, the mean value and the covariance of
the state variables) and of  the two-time correlation function.
These time-evolution equations will be derived directly from the
postulated linear stochastic equation with additive noise, rather
than from the time-evolution equation of the conditional probability
distribution, which is another possible route \cite{santamaria1}.

Onsager-Machlup's theory contains elements of purely mathematical
nature that must be clearly distinguished from the purely physical
assumptions about the behavior of physical systems. Among the
former, the most relevant is the concept of Ornstein-Uhlenbeck
process, i.e., a stationary Gaussian Markov stochastic process
\cite{ornsteinuhlenbeck,wanguhlenbeck}. This is the mathematical
model employed to state the main assumption of physical nature in
Onsager-Machlup's theory, which can then be stated by saying that
the spontaneous fluctuations of an equilibrium system can be
described as an Ornstein-Uhlenbeck process. Thus, after a brief
qualitative discussion in section \ref{II} on the assumed general
features of the thermodynamic conceptual framework that we have in
mind to describe the evolution of the system, section \ref{III}
summarizes the definition and the main properties of an
Ornstein-Uhlenbeck process as a mathematical model. In Section
\ref{IV} we employ these concepts to summarize Onsager's theory
(with this term we refer to Onsager-Machlup's theory plus Onsager's
own work on the properties of the linear relaxation equations of
irreversible thermodynamics \cite{onsager1,onsager2}). The
generality of the underlying mathematical model makes Onsager's
theory applicable to conditions not considered in the original work.
These extensions serve as the basis for its generalization to the
non-linear regime, which we present in Section \ref{V}. To simplify
the reading of this paper, in Section \ref{VI} we summarize the main
concepts and results that constitute this generalized canonical
theory and in Section \ref{VII} we provide an illustration of its
concrete use by reviewing its application to colloid dynamics. The
significance of the results of this particular application of the
generalized canonical theory is also discussed in the final section
\ref{VIII}.

\section{Non-equilibrium relaxation and entropy landscape}\label{II}

Let us consider a system whose macroscopic state is described by a
set of $C$ extensive variables $ a_i(t)$, $i=1,2,...,C$, which we
group as the components of a $C$-component (column) vector
$\textbf{a}(t)$. The fundamental postulate of the present
statistical thermodynamic theory of non-equilibrium processes is
that the dynamics of the state vector $\textbf{a}(t)$ constitutes a
\emph{multivariate stochastic process}, described by a deterministic
equation for its mean value ${\overline {\textbf a }}(t)$ and by a
linear stochastic equation with additive noise for the fluctuations
$\delta {\textbf a }(t)\equiv \textbf{a}(t)-{\overline {\textbf a
}}(t)$. The basic assumption is that the mean value ${\overline
{\textbf a }}(t)$ coincides with the macroscopically measured value,
and that its time evolution is described by an equation of the
general form

\begin{equation}
\frac{d{\overline {\textbf a }}(t)}{dt}= \mathcal{R}\left[{\overline
{\textbf a }}(t)   \right], \label{releq0}
\end{equation}
where the (generally non-linear and spatially non-local) functional
dependence  of the $C$-component vector $\mathcal{R}\left[{\overline
{\textbf a }}(t) \right]$ on ${\overline {\textbf a }}(t)$ includes
both, dissipative and mechanical (i.e., conservative) terms
\cite{keizer}.

Under some conditions, the relaxation equation above may admit
time-independent, or stationary, solutions denoted by ${\textbf
a}^{ss}$. This means that ${\textbf a}^{ss}$  solves the equation

\begin{equation}
\frac{d{\textbf a}^{ss}}{dt}= \mathcal{R}\left[{\overline {\textbf a
}}^{\ ss} \right]=0. \label{stationary0}
\end{equation}
Stationary states may result from the non-linear mathematical
structure of the relaxation equation, Eq. (\ref{releq0}). These are
``true" non-equilibrium stationary states, in the sense that their
stability requires the continuous input and output of energy,
matter, etc., thus involving intrinsically open systems. In our
present discussion we shall not refer to this kind of stationary
states. Instead, we refer to stationary states that correspond to
absolute or local maxima of the total entropy, implying conditions
for the thermodynamic equilibrium of the system within the
constraints imposed by its interactions with external fields and
reservoirs and within the internal constraints resulting from, and
sustained by, strong intermolecular interactions, such as in
meta-stable or in arrested states.

\begin{figure}[ht]
\includegraphics[scale=.4]{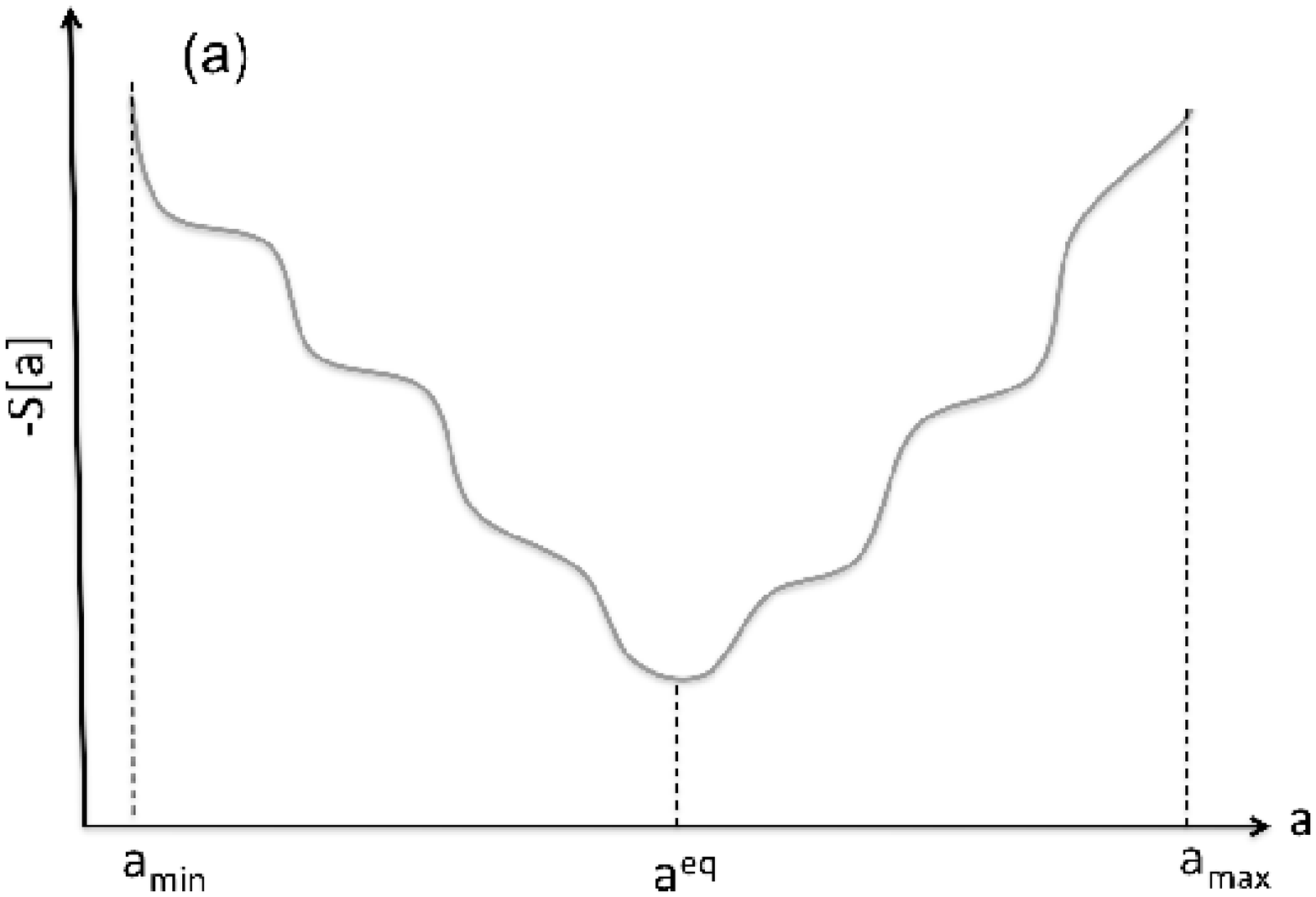}
\includegraphics[scale=.4]{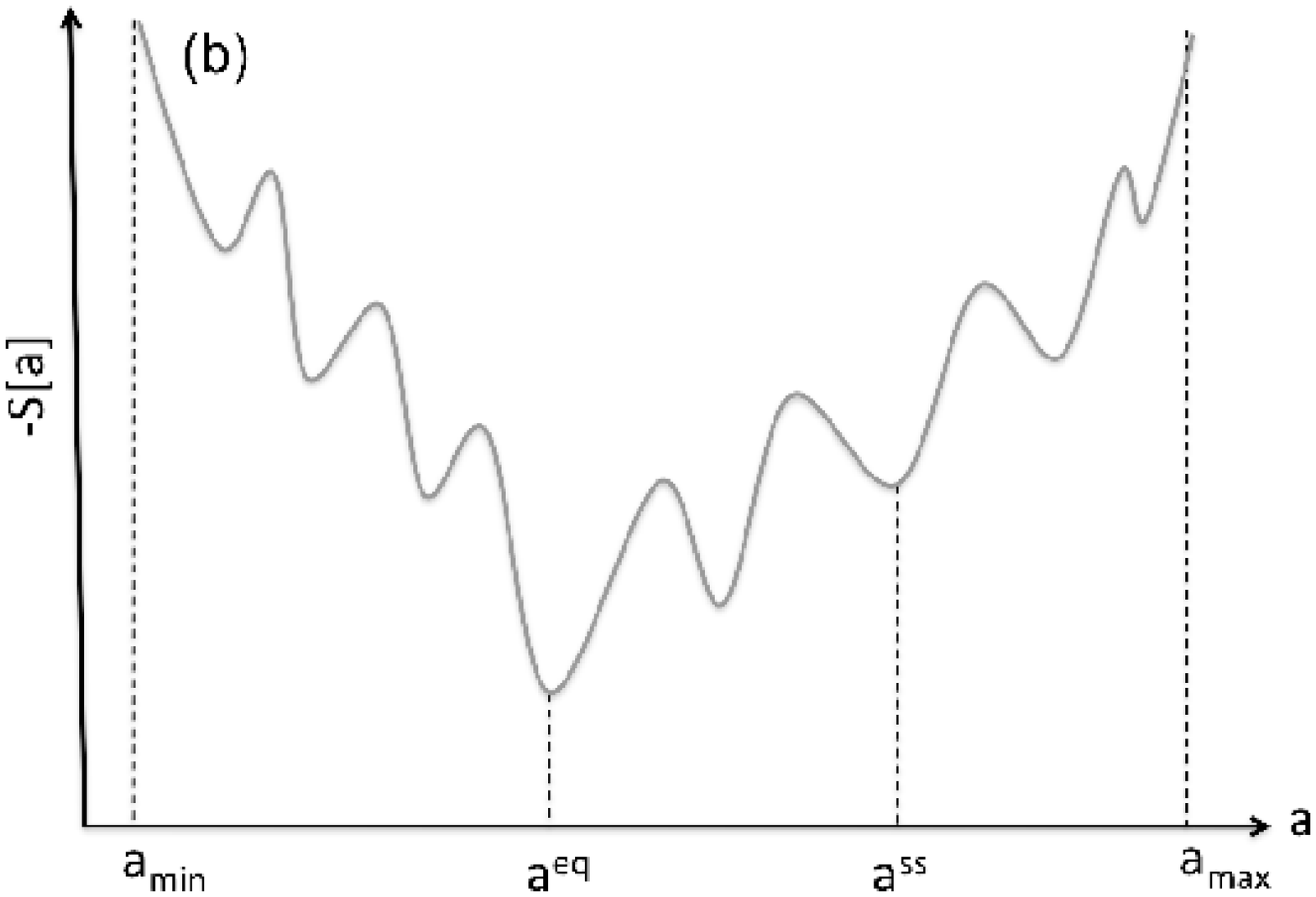}
\caption{Schematic illustration of the ``entropy" landscape of an
isolated system whose fundamental thermodynamic relation $S=S[a]$
involves a single macroscopic variable $a$. The isolation condition
constrains the possible states $a$ to a subset ${\mathcal T}$ of the
entire state space of the system; in the figure, this isolation
domain ${\mathcal T}$, is represented by the interval $a_{min}\leq a
\leq a_{max}$. The absolute entropy maximum is located at
$a=a^{eq}$. Fig. (a) illustrates the extreme case in which this
global entropy maximum is the only maximum, whereas Fig. (b)
illustrates the case in which there are many additional local
entropy maxima, one of them indicated by its location $a=a^{ss}$}.
\label{fig1}
\end{figure}

For a \emph{closed} system with fixed external constraints (i.e.,
with given time-independent external fields, including those of the
isolating and confining walls), there is one state that corresponds
to the absolute maximum of the function entropy $S=S[\textbf{a}]$
within the set ${\mathcal T}$ of macroscopic states $\textbf{a}$
consistent with those isolation conditions. According to the second
law of thermodynamics, such state, denoted by ${\textbf a}^{eq}$, is
the thermodynamically most stable equilibrium state. Fig. 1(a)
illustrates schematically the simplest scenario in terms of an
``entropy landscape" that exhibits a single and ``unquestionable"
entropy maximum. This global maximum ${\textbf a}^{eq}$ must then be
the only attractive stationary state of the non-linear relaxation
equation (\ref{releq0}), whose basin of attraction must then
coincide with the whole subspace ${\mathcal T}$ of states that are
consistent with the isolation condition. This means that if the
system is initially prepared in any state ${\textbf a
}^{0}\in{\mathcal T}$, its mean value ${\overline {\textbf a
}}^0(t)$ will evolve in time according to Eq. (\ref{releq0})
relaxing eventually, but surely, to the equilibrium state ${\textbf
a}^{eq}$.

Severe \emph{internal} constraints may, however, give rise to very
many additional equilibrium states, denoted generically as ${\textbf
a}^{ss}$, that correspond to local entropy maxima, as illustrated
schematically in Fig. 1(b). In contrast with the previous scenario,
in this case, depending on the location of the initial state
${\textbf a }^{0}\in{\mathcal T}$, the system will not necessarily
evolve towards the absolute thermodynamic equilibrium state $
{\textbf a }^{eq}$, since there are now many competing local entropy
maxima which also act locally (i.e., within its own basin of
attraction) as attractive stationary states representing
\emph{meta-stable} thermodynamic equilibrium states. In fact, each
of these meta-stable states, such as that represented by ${\textbf
a}^{ss}$ in Fig. 1(b), will actually be as stable as the ``most
stable" equilibrium state ${\textbf a}^{eq}$, as long as no external
(or internal!) perturbations drive the system, even momentarily, out
of the basin of attraction ${\mathcal T}({\textbf a }^{ss})$ of this
particular stationary state. We assume that the basins of attraction
of each pair of different stationary states in ${\mathcal T}$ are
disjoint, i.e., that ${\mathcal T}({\textbf a
}_1^{ss})\bigcap{\mathcal T}({\textbf a }_2^{ss})=\emptyset$ for any
pair ${\textbf a }_1^{ss}$ and ${\textbf a }_2^{ss}$, and that the
union of the basins of attraction of all these stationary states,
including ${\textbf a}^{eq}$, equals the state space ${\mathcal T}$
of the isolated system, $\bigcup _{{\textbf a }^{ss}}{\mathcal
T}({\textbf a }^{ss})={\mathcal T}$.

In the previous discussion we have assumed that the entropy
landscape in Fig. 1(b) remains constant in time so that if the
system was prepared in the basin of attraction of a given stationary
state, it will be trapped in this basin, eventually relaxing to the
corresponding stationary state ${\textbf a}^{ss}$, in which it will
remain indefinitely afterwards. This means that all the solutions of
the relaxation equation in Eq. (\ref{linearreleq}) consist of
trajectories confined to the basin of attraction of one specific
stationary state. In reality, it is impossible that such situation
can be sustained indefinitely, since there are many possible
mechanisms that allow the trajectory ${\overline {\textbf a }}^0(t)$
to explore states outside the basin of attraction in which the
system was initially prepared. If such excursion occurs to the basin
of a stationary state with a still larger local entropy maximum, the
second law of thermodynamics dictates that the system will proceed
to the stationary state of the new basin of attraction. If this
process occurs repeatedly, the system will eventually reach the most
stable equilibrium state ${\textbf a}^{eq}$, thus restoring to some
extent the irreversible process described in the simpler conditions
illustrated in Fig. 1(a), but with a probably much slower dynamics
because the transitions from basin to basin may involve activation
barriers that must await for the occurrence of adequate spontaneous
fluctuations. In this process, the lifetime of these instantly
stable stationary states may exceed the experimental timescales,
thus giving rise to meta-stability conditions, and/or to conditions
in which the system may appear to be dynamically arrested.

A fundamental assumption of the present theory will be that the
actual relaxation of a system under the more complex conditions
illustrated in Fig. 1(b) can be approximated by a sequence
${\overline {\textbf a }}^0(t_\alpha) = {\textbf a}_\alpha^{ss}$
($\alpha =$ 0, 1, 2,...) of locally stationary states ${\textbf
a}_\alpha^{ss}$ generated by the recurrence relation ${\textbf
a}_{\alpha+1}^{ss} = {\textbf a}_\alpha^{ss} + \mathcal{R}\left[
{\textbf a }_\alpha^{\ ss} \right](t_{\alpha+1}-t_{\alpha})$,
starting with ${\textbf a}_{0}^{ss}={\textbf a}^{0}$. Although these
locally stationary states ${\textbf a}^{ss}$ are not stable
equilibrium states, we will assume that their properties may be
described by quite similar fundamental principles as the most stable
equilibrium state ${\textbf a}^{eq}$. Clearly, the conceptual
framework and the approximations employed in the theoretical
modeling of the dynamics of ${\overline {\textbf a }}^0(t)$ must
reflect the fundamental difference between the two qualitatively
different scenarios illustrated in Fig. 1. Onsager's linear
irreversible thermodynamic theory of fluctuations
\cite{onsager1,onsager2,onsagermachlup1, onsagermachlup2} has in
mind the simpler scenario illustrated by the entropy landscape of
Fig. 1(a). We are interested, however, in generalizing Onsager's
theory to the conditions represented schematically by the more
complex scenario in Fig. 1(b). The fundamental postulate of
Onsager's theory is that the macroscopic dynamics of the system is
not described by a deterministic equation for the state vector
$\textbf{a}(t)$. Instead, it is postulated that the macroscopic
state of the system is described by a statistical physical ensemble
whose mathematical representation involves the assumption that
$\textbf{a}(t)$ constitutes a multivariate stochastic process, and
more specifically, a stochastic process referred to as an
Ornstein-Uhlenbeck process. Thus, before explaining the main
\emph{physical} assumptions made in Onsager's theory, let us review
the \emph{purely mathematical} framework of stochastic processes, in
which such physical assumptions can be stated most economically.

\section{Ornstein-Uhlenbeck process, a reference mathematical
model}\label{III}

In this section we summarize the main concepts that define an
Ornstein-Uhlenbeck process
\cite{keizer,ornsteinuhlenbeck,wanguhlenbeck}. As a mathematical
object, a stochastic process $\textbf{a}(t)$ is defined in terms of
the joint probability density
$W_m(\textbf{a}_1,t_1;\textbf{a}_2,t_2;...;\textbf{a}_m,t_m)$ for
the state vector $\textbf{a}(t)$ to have a value in the interval
$\textbf{a}_i \le \textbf{a}(t_i) \le \textbf{a}_i + d \textbf{a}_i$
for $i=1,2,...,m$. We say that this stochastic process is fully
determined if we know the probability densities for all possible
positive integer values of $m$ and all possible sets of times
$(t_1,t_2,...,t_m)$. If the stochastic process is \emph{Markovian},
however, a great simplification occurs, since in this case all these
probability densities can be written in terms of only
$W_2(\textbf{a}_1,t_1;\textbf{a}_2,t_2)$. This probability density
can be written as
$W_2(\textbf{a}_1,t_1;\textbf{a}_2,t_2)=W_1(\textbf{a}_1,t_1)P_2(\textbf{a}_1,t_1\mid\textbf{a}_2,t_2)$
where $P_2(\textbf{a}_1,t_1\mid\textbf{a}_2,t_2)$ is the conditional
probability density that $\textbf{a}(t_2)$ has a value in the
interval $\textbf{a}_2 \le \textbf{a}(t_2) \le \textbf{a}_2 + d
\textbf{a}_2$ provided that for sure $\textbf{a}(t_1)=\textbf{a}_1$.

A stochastic process is said to be \emph{stationary} if all its
probability densities are time-translational invariant, i.e., if for
all real values of $s$ we have that
$W_m(\textbf{a}_1+s,t_1;\textbf{a}_2,t_2+s;...;\textbf{a}_m,t_m+s)=
W_m(\textbf{a}_1,t_1;\textbf{a}_2,t_2;...;\textbf{a}_m,t_m)$. Thus,
if in addition to being Markovian the stochastic process is also
stationary, then $W_1(\textbf{a}_1,t_1)=W(\textbf{a}_1)$ and
$P_2(\textbf{a}_1,t_1\mid\textbf{a}_2,t_2)=
P(\textbf{a}_1\mid\textbf{a}_2,t_2-t_1)$. Assuming that
$\lim_{t\to\infty}P(\textbf{a}_1\mid\textbf{a}_2,t)=W(\textbf{a}_2)$,
then we have that a stationary Markov process is determined solely
by the conditional probability density
$P(\textbf{a}_1\mid\textbf{a}_2,t_2-t_1)$.

The knowledge of this probability density is equivalent to the
knowledge of all its moments. A final great simplification occurs
when we assume that the stationary Markov process is, additionally,
Gaussian, i.e., such that all the moments of
$P(\textbf{a}_0\mid\textbf{a},t)$ can be written in terms of only
its two lowest-order moments as

\begin{equation}
P({\textbf a }^0\mid{\textbf a },t)= [(2\pi)^C \det{\sigma(t)}]
e^{-[({\textbf a }-{\overline{\textbf a}}^0(t))^{\dagger}\circ
\sigma ^{-1}(t)\circ ({\textbf a }-{\overline {\textbf a
}}^0(t))]/2}, \label{gaussianpat}
\end{equation}
where the dagger means transpose, the circle ``$\circ$" means matrix
product, and where the conditional mean value
$\overline{\textbf{a}}^0(t)$ and the covariance matrix $\sigma(t)$
of the fluctuations $\delta {\textbf a }(t)\equiv {\textbf a }(t)-
{\overline{\textbf a}}^0(t)$ are defined, respectively, as

\begin{equation}
\overline{\textbf{a}}^0(t) \equiv \int
\textbf{a}P(\textbf{a}_0\mid\textbf{a},t)d\textbf{a} \label{defmean}
\end{equation}
and

\begin{equation} \sigma(t) \equiv
\overline{\delta\textbf{a}(t)\delta\textbf{a}^{\dagger}(t)} \equiv
\int (\textbf{a}-\overline{\textbf {a}}^0(t))
(\textbf{a}-\overline{\textbf {a}}^0(t))^{\dagger}
P(\textbf{a}_0\mid\textbf{a},t)d\textbf{a}.\label{defcovariance}
\end{equation}
We notice that in the long-time limit,
$P(\textbf{a}_0\mid\textbf{a},t)$ attains its stationary value $
W({\textbf a })= [(2\pi)^C \det{\sigma^{ss}}] e^{-[({\textbf a
}-{\overline {\textbf a }}^{ss})^{\dagger}\circ \sigma ^{ss-1}\circ
({\textbf a }-{\overline {\textbf a }}^{ss})]/2}$, with ${\overline
{\textbf a }}^{ss}\equiv {\overline{\textbf a}}^0(t\to\infty)$ and
$\sigma^{ss}\equiv \sigma(t\to\infty)$.

A stochastic process that is stationary, Gaussian, and Markov, like
the one just discussed, has a proper name: it is referred to as an
\emph{Ornstein-Uhlenbeck} process \cite{ornsteinuhlenbeck,
wanguhlenbeck}. From a mathematical point of view this is a most
attractive and useful mathematical model, since its full
determination is reduced to the determination of two objects,
namely, the mean value $\overline{\textbf {a}}^0(t)$ and the
covariance $\sigma(t)$, whose determination then becomes a central
issue. This issue, however, is partially settled when one
demonstrates that each of these two properties satisfies a
deterministic relaxation equation whose structure is also determined
by the model itself, as we now discuss.

\subsection{Relaxation equations for $\overline{\textbf {a}}^0(t)$ and
$\sigma(t)$}\label{III.1}

Besides the definition given above, the Ornstein-Uhlenbeck
stochastic process can be given other alternative but mathematically
equivalent definitions \cite{keizer}. The most relevant for our
present purpose defines the stochastic process $ {\textbf a }(t)$ as
the stationary ensemble constituted by all the possible
``realizations" of the stochastic process, which correspond to all
the possible solutions of a linear stochastic differential equation
with additive noise for the fluctuations $\delta {\textbf a
}(t)\equiv {\textbf a }(t)-{\overline {\textbf a }}^{ss}$. Such
stochastic equation has the following general structure

\begin{equation}
\frac{d \delta {\textbf a }(t)}{dt}=   \mathcal{H} \circ \delta
{\textbf a }(t) +{\textbf f}(t), \label{releq0001}
\end{equation}
with $\mathcal{H}$ being a $C\times C$ relaxation matrix and with
the $C$-component vector ${\textbf f}(t)$ being a ``white noise",
i.e., a stationary and Gaussian stochastic process which is,
however, not Markovian, but $\delta$-correlated, i.e., such that
$<\textbf{f}(t)\textbf{f}^{\dagger}(t')>= \gamma 2\delta (t-t')$
(with $``<\cdot\cdot\cdot>"$ indicating the average over all the
possible realizations of the noise $\textbf{f}(t)$). In addition, it
is assumed that $<{\textbf f}(t)>={\textbf 0}$ and $<{\textbf
f}(t)\delta {\textbf a }^{\dagger}(0)>=0$. The stochastic process
$\delta {\textbf a }(t)$ thus defined can be shown to be Gaussian,
Markov, and stationary, i.e., to be an Ornstein-Uhlenbeck process.

The main reason one might prefer to define a stochastic process in
terms of a linear stochastic equation with additive noise, is that
one can derive the deterministic time evolution equations for
$\overline{\textbf {a}}^0(t)$ and $\sigma(t)$ much more directly
from such stochastic equation than from the definition of these
properties in Eqs. (\ref{defmean}) and (\ref{defcovariance}), which
require the previous derivation of the time-evolution equation of
the full probability distribution function
$P(\textbf{a}_0\mid\textbf{a},t)$. To illustrate this, let us notice
that the solution of Eq. (\ref{releq0001}) can be written as

\begin{equation}
\delta {\textbf a }(t) = e^{\mathcal{H}t} \circ \delta {\textbf a
}(0) + \int _0^t e^{\mathcal{H}(t-\tau)} \circ  \textbf{f}(\tau)
d\tau\label{soloueq}
\end{equation}
with $\delta {\textbf a }(0)={\textbf a }^0-{\textbf a}^{ss}$.
Notice also that the mean value $\overline{\delta\textbf
{a}(t)}^0=\overline{\textbf {a}}^0(t)-{\textbf{a}}^{ss}$ is the
average of $\delta\textbf{a}(t)$ over the realizations
$\textbf{f}(t)$ but not over the initial conditions $\delta {\textbf
a }(0)$, i.e., $\overline{\delta\textbf {a}(t)}^0 = < \delta{\textbf
a }(t)>$ with the initial condition $\overline{\delta\textbf{a}(0)}
= {\textbf a }^0-{\textbf a}^{ss}$. Thus, taking the average of  Eq.
(\ref{soloueq}) over the realizations $\textbf{f}(t)$, it follows
that the $\overline{\textbf {a}}^0(t)$ is given by

\begin{equation}
\overline{\textbf {a}}^0(t)= {\textbf a}^{ss} + e^{\mathcal{H}t}
\circ [{\textbf a }^0-{\textbf a}^{ss}].  \label{meansoloueq}
\end{equation}
As long as the real part of all the eigenvalues of the relaxation
matrix $\mathcal{H}$ remain negative, this expression interpolates
$\overline{\textbf {a}}^0(t)$ between its arbitrary initial
condition ${\textbf a }^0$ and its asymptotic stationary value
${\textbf a}^{ss}$. Clearly, this expression for $\overline{\textbf
{a}}^0(t)$ is the solution of the ordinary differential equation

\begin{equation}
\frac{d \Delta \overline{{\textbf a }}(t)}{dt}=   \mathcal{H} \circ
\Delta \overline{{\textbf a }}(t), \label{releqmean}
\end{equation}
where $\Delta \overline{{\textbf a }}(t)\equiv \overline{\textbf
{a}}^0(t)-{\textbf a}^{ss}$, with initial condition $\Delta
\overline{{\textbf a }}(0)\equiv \textbf{a}^0-{\textbf a}^{ss} $.
Thus, we conclude that in the Ornstein-Uhlenbeck process the
stochastic equation for the fluctuations $\delta {\textbf a }(t)$
coincides with the equation that governs the decay of the deviation
$\Delta \overline{{\textbf a }}(t)$ of the conditional mean value
$\overline{\textbf {a}}^0(t)$ except for the additive white noise
term ${\textbf f}(t)$ in Eq. (\ref{releq0001}). We also notice that
the previous relaxation equation for $\overline{\textbf {a}}^0(t)$,
as well as its explicit solution in Eq. (\ref{meansoloueq}), only
depend on the relaxation matrix $\mathcal{H}$ for given initial
condition $\textbf{a}^0$.

The relaxation equation for the covariance $\sigma(t)$ can be
derived in a similar manner. Thus, by multiplying Eq.
(\ref{soloueq}) on the right by its transpose, and averaging the
resulting expression both, over the realizations of $\textbf{f}(t)$
and over the initial values $\delta {\textbf a }(0)$, it is not
difficult to derive the following expression for $\sigma(t)$,

\begin{equation}
\sigma(t) =  e^{\mathcal{H}t} \circ [\sigma^0 + 2\int _0^t
e^{-\mathcal{H}\tau} \circ \gamma \circ  e^{-\mathcal{H}^{
\dagger}\tau} d\tau]  \circ e^{\mathcal{H}^{\dagger }t},
\label{sigmadt}
\end{equation}
where $\sigma^0$ is the given initial value of $\sigma(t)$. By
differentiating this expression with respect to time, $\sigma(t)$ is
seen to satisfy the differential equation

\begin{equation}
\frac{d\sigma(t)}{dt} = \mathcal{H}\circ \sigma (t) + \sigma (t)
\circ \mathcal{H}^{\dagger } + 2\gamma \label{sigmadteq}
\end{equation}
with initial condition $\sigma(0)=\sigma^0$.

\subsection{The fluctuation-dissipation theorem}\label{III.2}

The formal solution for $\sigma(t)$ in Eq. (\ref{sigmadt}) can be
written as $\sigma(t) =  e^{\mathcal{H}t} \circ \sigma^0 \circ
e^{\mathcal{H}^{\dagger }t} + 2\int _0^t e^{\mathcal{H}s} \circ
\gamma \circ e^{\mathcal{H}^{ \dagger}s} ds $ which, in the limit $t
\to \infty$, reads $\sigma^{ss} = 2\int _0^\infty e^{\mathcal{H}s}
\circ \gamma \circ e^{\mathcal{H}^{ \dagger}s} ds $. Subtracting
this expression from the previous one, and using the change of
variable $\tau=s-t$, Eq. (\ref{sigmadt}) may be given the following
alternative expression

\begin{equation}
\sigma(t)= \sigma^{ss} + e^{\mathcal{H}t}\circ
[\sigma^{0}-\sigma^{ss}]\circ e^{\mathcal{H}^{\dagger }t},
\label{solsigmadt}
\end{equation}
which interpolates the covariance $\sigma(t)$ between its given
initial value $\sigma^0$ and its asymptotic stationary value
$\sigma^{ss}$. Taking the time derivative of this expression and
comparing it with Eq. (\ref{sigmadteq}), finally leads to the
so-called fluctuation-dissipation theorem,

\begin{equation}
\mathcal{H}\circ \sigma^{ss} + \sigma^{ss} \circ
\mathcal{H}^{\dagger } +2\gamma =0. \label{fdtheorem0}
\end{equation}

This relationship between the matrices $\mathcal{H}$, $\sigma^{ss}$,
and $\gamma$ constitutes a necessary and sufficient condition for
the process to be stationary, and also follows directly from Eq.
(\ref{sigmadteq}) provided that the long-time stationary solution
$\sigma^{ss}$ exists. Besides guaranteeing stationarity, this
relationship also sets stringent conditions, of purely mathematical
nature, on the structure of the relaxation matrix $\mathcal{H}$. To
see this let us define the matrix $\mathcal{L}$ as the product
\begin{equation}
\mathcal{L}\equiv - \mathcal{H}\circ \sigma^{ss}.  \label{matrixl}
\end{equation}
Then, the fluctuation-dissipation relationship in Eq.
(\ref{fdtheorem0}) can be written as
\begin{equation}
\gamma = \frac{\mathcal{L}+\mathcal{L}^\dagger}{2}\equiv
\mathcal{L}_s. \label{fdtheoremlsim}
\end{equation}
This equation can be viewed as a condition that determines the
matrix $\gamma$ (a measure of the time-correlation function of the
additive white noise $\textbf{f}(t)$) in terms of the symmetric part
$\mathcal{L}_s$ of the matrix $\mathcal{L}$ defined, in its turn, in
Eq. (\ref{matrixl}) in terms of the relaxation matrix $\mathcal{H}$
and the stationary covariance $\sigma^{ss}$.

The same equation can also be viewed as a condition of the general
structure of the relaxation matrix $\mathcal{H}$, which then must be
such that it can be factorized as

\begin{equation}
\mathcal{H} = - \mathcal{L}\circ \sigma^{ss-1},  \label{matrixh}
\end{equation}
with the matrix $\mathcal{L}$ being identical to the symmetric
matrix $\gamma$ plus, at most, an antisymmetric matrix
$\mathcal{L}_a \equiv (\mathcal{L}-\mathcal{L}^\dagger)/2$. Thus, we
conclude that the fluctuation-dissipation relationship dictates that
the relaxation matrix $\mathcal{H}$ is not completely arbitrary, but
must comply with the rather rigid format of Eq. (\ref{matrixh}). For
example, the most general structure of the linear relaxation
equation for the mean value $\Delta\overline{{\textbf a }}(t)$ in
Eq. (\ref{releqmean}) must be of the form

\begin{equation}
\frac{d \Delta \overline{{\textbf a }}(t)}{dt}=   - \mathcal{L}\circ
\sigma^{ss-1} \circ \Delta \overline{{\textbf a }}(t),
\label{releqmean2}
\end{equation}
a format which, in the case that the stationary state
$\textbf{a}^{ss}$ is the thermodynamic equilibrium state
$\textbf{a}^{eq}$, will soon be identified with the canonical form
of the relaxation equations of linear irreversible thermodynamics.

\subsection{Time-dependent correlation function
$C(t',t)$}\label{III.3}

Another important property of the stochastic process $\textbf{a}(t)$
is its two-time correlation matrix $C(t',t)\equiv
\overline{\delta\textbf{a}(t') \delta\textbf{a}^\dagger(t) }$ where
the over-line refers to an average over all the possible
realizations of the stationary noise $\textbf{f}(t)$ plus the
average over the distribution of the initial values
$\delta\textbf{a}(0)$. We may evaluate this property by using again
Eq. (\ref{soloueq}) evaluated at $t'$ and at $t$ to form the product
$\delta\textbf{a}(t') \delta\textbf{a}^\dagger(t)$. Taking the
average of this product, we may use the fact that
$\overline{\textbf{f}(t')\textbf{f}^{\dagger}(t)}= \gamma 2\delta
(t'-t)$ and $\overline{{\textbf f}(t)\delta {\textbf a
}^{\dagger}(0)}=0$ to show that $C(t',t)$ may be written, assuming
that $t'>t$, as
\begin{equation}
C(t',t)=e^{\mathcal{H}(t'-t)}\circ\sigma(t),  \label{ctpt}
\end{equation}
where the expression for $\sigma(t)$ in Eq. (\ref{sigmadt}) was
employed.

Defining $t'=t+\tau$, one can also write $C(t',t)$ as $C(t+\tau,t)=
\overline{\delta\textbf{a}(t+\tau) \delta\textbf{a}^\dagger(t)}
\equiv C_t(\tau)$ to describe the decay of the correlations with the
\emph{correlation time} $\tau$ when the system has evolved during a
\emph{relaxation time} $t$ after it was prepared in an initial state
represented by the distribution of initial values
$\delta\textbf{a}(0)$ with mean value
$\overline{\delta\textbf{a}(0)}=\textbf{a}^0-\textbf{a}^{ss}$ and
covariance $\sigma^0$. One can also derive the time evolution of $
C_t(\tau)$ by writing $t+\tau$ as the argument of
$\delta\textbf{a}(t)$ in Eq. (\ref{soloueq}), whose derivative with
respect to $\tau$ for fixed $t$ can be shown to be given by
\begin{equation}
\frac{d \delta {\textbf a }(t+\tau)}{d\tau}=   \mathcal{H} \circ
\delta {\textbf a }(t+\tau) +{\textbf f}(t+\tau). \label{otravez}
\end{equation}
Multiplying this result on the right by the transpose of
$\delta\textbf{a}(t)$ and averaging over the realizations of
$\textbf{f}(t+\tau)$, one is led to the following relaxation
equation for $C(t+\tau,t)$,
\begin{equation}
\frac{dC(t+\tau,t)}{d\tau}=\mathcal{H}\circ C(t+\tau,t),
\label{dcttp}
\end{equation}
with initial condition $C(t+0,t)= \sigma(t)$. The solution of this
equation is again given by Eq. (\ref{ctpt}), now written as
\begin{equation}
C(t+\tau,t)=e^{\mathcal{H}\tau}\circ\sigma(t) \ \ \ \ \ \ \
(\textrm{for}\ \tau\ge0).  \label{cttp}
\end{equation}

Let us notice that for sufficiently long relaxation times $t$ after
the initial condition set at $t=0$, the process recovers its
stationary condition, in which $\sigma(t\to\infty) = \sigma^{ss}$
and $C(t+\tau,t)=C(\tau,0)\equiv C^{ss}(\tau)$, with

\begin{equation}
C^{ss}(\tau) \equiv
\lim_{t\to\infty}C(t+\tau,t)=e^{\mathcal{H}\tau}\circ\sigma^{ss} \ \
\ \ \ \ \ (\textrm{for}\ \tau\ge0).  \label{cttpp}
\end{equation}

In summary, the Ornstein-Uhlenbeck stochastic process is fully
determined by the time evolution of the mean value
$\overline{\textbf {a}}^0(t)$ and the covariance $\sigma(t)$,
governed by the deterministic equations (\ref{releqmean}) and
(\ref{sigmadteq}). The solution of these equations, given in Eqs.
(\ref{meansoloueq}) and (\ref{solsigmadt}), is a simple exponential
interpolation between the given initial state
($\textbf{a}^0,\sigma^0$) and the final stationary state
($\textbf{a}^{ss},\sigma^{ss}$) with relaxation times given by the
eigenvalues of the relaxation matrix $\mathcal{H}$. This matrix,
however, is not completely arbitrary, but must conform to the format
in Eq. (\ref{matrixh}). Nevertheless, this matrix becomes the kernel
of the model, since it is the main ingredient in terms of which all
the properties of the process are determined. This includes, besides
$\overline{\textbf {a}}^0(t)$ and $\sigma(t)$, the time-dependent
correlation function $C(t',t)$ just discussed, and written
explicitly in terms of $\mathcal{H}$ in Eq. (\ref{ctpt}) or
(\ref{cttp}).

\section{Onsager's theory of equilibrium thermal fluctuations}\label{IV}

The mathematical infrastructure just reviewed can now be employed to
described specific physical phenomena in a  very efficient manner.
This approach was inaugurated by Langevin \cite{langevin} with his
celebrated equation for the thermal fluctuations of the velocity
$\textbf{V}(t)$ of a Brownian particle. In reality his work
triggered the development of the mathematical field of stochastic
processes \cite{wax}, from which we borrowed the concepts summarized
in the previous section. Today, however, we may state his theory
most efficiently as the assumption that the cartesian vector
$\textbf{V}(t)$ constitutes an Ornstein-Uhlenbeck process specified
by the diagonal relaxation matrix
$\mathcal{H}_{ij}=-(\zeta/M)\delta_{ij}$ (with $i,j=1,2,3$, and with
$\zeta$ and $M$ being the friction coefficient and the mass of the
Brownian particle). From this assumption, all the results of the
previous section apply with $\textbf{a}(t)$ replaced by
$\textbf{V}(t)$, and with this specific value of $\mathcal{H}$.

With this identification of the relaxation matrix $\mathcal{H}$, the
textbook presentations of Langevin's theory of Brownian motion
\cite{mcquarrie} then become just a review of the physical meaning,
worded in the specific context of Brownian motion, of the
mathematical properties of the underlying Ornstein-Uhlenbeck
mathematical model summarized in the previous section. For example,
according to Eq. (\ref{releq0001}), and as Langevin puts it in his
original work \cite{langevin}, the thermal fluctuations
${\textbf{V}}(t)$ are then governed by the stochastic version of
this linearized Newton's equation, namely, $M(d\textbf{V}(t)/dt)=
-\zeta\textbf{V}(t)+\textbf{f}(t)$. The random force $\textbf{f}(t)$
is then assumed to be ``indifferently positive and negative and with
a magnitude such that it maintains the agitation of the particle,
which the viscous resistance would stop without it", a physically
intuitive manner employed by Langevin to express his assumption that
$\textbf{f}(t)$ is a \emph{stationary} delta correlated noise.

It is then instructive to recall Langevin's arguments leading to the
identification of the relaxation matrix $\mathcal{H}$ above. These
are extremely simple: the conditional mean velocity
$\overline{\textbf{V}}(t)$ of a colloidal particle obeys the same
equation that governs the motion of a macroscopic particle under the
influence of the frictional resistance
$\textbf{R}[\overline{\textbf{V}}(t)]$ of the solvent, i.e.,
Newton's equation, $M(d\overline{\textbf{V}}(t)/dt)=
\textbf{R}[\overline{\textbf{V}}(t)]$. The functional dependence of
the resistance $\textbf{R}[\overline{\textbf{V}}(t)]$ on the
velocity ${\textbf{V}}(t)$, referred to as the ``constitutive"
relation, is in general non-linear and poorly understood, but its
linear approximation
$\textbf{R}[\overline{\textbf{V}}(t)]\approx-\zeta\overline{\textbf{V}}(t)$
leads to the simplest form of the relaxation equation for
$\overline{\textbf{V}}(t)$, namely,
$M(d\overline{\textbf{V}}(t)/dt)= -\zeta\overline{\textbf{V}}(t)$.
Comparing with Eq. (\ref{releqmean}), one immediately identifies the
relaxation matrix to be $\mathcal{H}_{ij}=-(\zeta/M)\delta_{ij}$.

\subsection{Linear laws and linear regime of irreversible
thermodynamics}\label{IV.1}

Onsager's theory of equilibrium thermal fluctuations is, to a large
extent, an extension of Langevin's theory, in which the vector
$\textbf{V}(t)$ is extended to be a generic state vector
$\textbf{a}(t)$. Just like in Langevin's theory, the first main
postulate of Onsager's theory is that the macroscopically observable
state variables correspond to the mean value ${\overline {\textbf a
}}(t)$ of a stochastic process whose time evolution is governed by a
nonlinear relaxation equation that we write in the general form of
Eq. (\ref{releq0}), namely,

\begin{equation}
\frac{d{\overline{\textbf a}}(t)}{dt}= \mathcal{R}\left[{\overline
{\textbf a }}(t)   \right]. \label{releq0repet}
\end{equation}
This general non-linear equation is assumed to represent, for
example, the hydrodynamic transport equations or Fick's diffusion
equation, as particular cases \cite{keizer,degrootmazur}. The
generally nonlinear dependence of the ``flux"
$\mathcal{R}_i\left[{\overline{\textbf a}}(t) \right]$ (the time
rate of change of the extensive variable $ {\overline a }_i(t)$) on
the state vector $\overline{\textbf{a}} (t)$ is referred to as the
\emph{constitutive relation}, a relation that turns Eq.
(\ref{releq0repet}) into a closed equation for
$\overline{\textbf{a}} (t)$.

Although the explicit form of this relation is not known in general,
in the so-called linear regime, discussed below, the resulting
relaxation equation must conform exactly to a rather strict format.
According to the point of view adopted in the conventional account
of Onsager's theory \cite{onsager1, onsager2,keizer,degrootmazur},
the format of the relaxation equations in the linear regime is a
direct consequence of one of the most fundamental \emph{physical}
principles of non-equilibrium irreversible thermodynamics, namely,
the phenomenological so-called \emph{linear laws}
\cite{degrootmazur}. This principle states that the ``flux"
$\mathcal{R}_i\left[{\overline{\textbf a}}(t) \right]$ is
proportional to the $C$-component vector of ``thermodynamic forces",
$\Delta {\textbf F }(t)\equiv {\textbf F }[ {\overline {\textbf
a}}(t)]- {\textbf F }^{eq}$, whose components describe the
instantaneous imbalance of the conjugate intensive variables $  F_j[
{\overline{\textbf a}}(t)] \equiv k_B^{-1} \left(
\partial S[ {\textbf a}] /
\partial a_j \right)_ {{\textbf a}=\overline{\textbf a}(t)}$ with
respect to their equilibrium value $ F _j^{eq}$. Thus, the linear
laws state that
\begin{equation}
\mathcal{R}\left[{\overline{\textbf a}}(t)
\right]=-\mathcal{L}^*\left[{\overline{\textbf a}}(t) \right]\circ
\Delta {\textbf F }(t) \label{reqldf}
\end{equation}
with the proportionality matrix $\mathcal{L}^*\left[{\overline
{\textbf a }}(t) \right]$ being referred to as the matrix of
Onsager's ``kinetic coefficients". We assume that this matrix is
known in principle from phenomenological considerations, as in
Fourier's heat conduction or in Fick's diffusion laws. The star on
$\mathcal{L}^*\left[{\textbf a }\right]$ is as reminder of the
phenomenological nature of this property and of its dependence on
the state vector \textbf{a}. Clearly, this principle requires the
existence of the state function entropy $S=S[ {\textbf a}]$ which,
for given conditions of isolation and given fixed external fields on
the system, has its maximum at a particular state $ {\textbf
a}^{eq}$, as prescribed by the second law of thermodynamics. Let us
stress that the entropy $S$ refers to the entropy of the closed
system, in which case $ F _i^{eq}=0$; if the system is in contact
with thermal and/or chemical reservoirs, the corresponding $ F
_i^{eq}$ may, however, differ from zero.

The linear laws described by Eq. (\ref{reqldf}) do not necessarily
imply that the resulting relaxation equation for the macroscopic
variables ${\overline{\textbf a}}^0(t)$ will in general be linear.
Onsager's theory, however, is restricted to the so-called
\emph{linear regime} around the thermodynamic equilibrium state
$\textbf{a}^{eq}$. Thus, imagine that the system is prepared in an
initial state $ \textbf{a}_0$ that lies in the basin of attraction
$\mathcal{T}({\textbf a }^{eq})$ of the equilibrium state. Then,
after some time the trajectory $\overline {\textbf{a}}(t)$ will be
sufficiently close to ${\textbf a}^{eq}$ that Eq.
(\ref{releq0repet}) can be linearized in the difference $\Delta
{\overline{\textbf a}}(t) \equiv {\overline{\textbf a}}(t)- {\textbf
a}^{eq}$, to read

\begin{equation}
\frac{d\Delta  {\overline{\textbf a}}(t)}{dt} =
\mathcal{H^*}[\textbf{a}^{eq}]\circ \Delta {\overline {\textbf
a}}(t), \label{linearreleqh}
\end{equation}
with the elements $\mathcal{H^*}_{ij}[\textbf{a}^{eq}]$ of the
$C\times C$ relaxation matrix $\mathcal{H^*}[\textbf{a}^{eq}]$ being
$ \mathcal{H^*}_{ij}[\textbf{a}^{eq}]= \left(\partial
{\mathcal{R}_i[\textbf{a}]}/\partial a_j
\right)_{\textbf{a}=\textbf{a}^{eq}}$.

Of course, if we now assume that the fluxes
$\mathcal{R}\left[{\overline{\textbf a}}(t) \right]$ are indeed
given by Eq. (\ref{reqldf}), then we immediately see that the matrix
$\mathcal{H^*}[\textbf{a}^{eq}]$ is given by
\begin{equation}
\mathcal{H^*} [{\textbf a}^{eq}]= -\mathcal{L}^* [{\textbf a}^{eq}]
\circ \mathcal{E}\left[{\textbf a }^{\ eq}\right].
\label{relaxmatrixh}
\end{equation}
where the $C \times C$ matrix $\mathcal{E}\left[ {\textbf a
}\right]$ is defined as
\begin{equation}
\mathcal{E}_{ij}[{\textbf a }] \equiv -\left( \frac{\partial
F_i[{\textbf a }]}{\partial a_j} \right)=-\frac{1}{k_B}\left(
\frac{\partial^2 S[{\textbf a }]}{\partial a_i\partial a_j} \right),
\label{matrixE}
\end{equation}
and with $\mathcal{E}\left[{\textbf a}^{eq}\right]$ being this
matrix evaluated at the equilibrium state $  {\textbf a }^{eq}$.
Thus, in the neighborhood of the equilibrium state  linear
irreversible thermodynamics provides a very specific structure of
the linearized relaxation equation in Eq. (\ref{linearreleqh}),
namely,
\begin{equation}
\frac{d\Delta  {\overline{\textbf a}}(t)}{dt} = -\mathcal{L}^*
[{\textbf a}^{eq}] \circ \mathcal{E}\left[{\textbf a }^{\
eq}\right]\circ \Delta {\overline{\textbf a}}(t).
\label{linearreleq}
\end{equation}

The regime in which the relaxation of $\overline {\textbf{a}}(t)$ is
described by this linear relaxation equation is referred to as the
\emph{linear regime} around ${\textbf a}^{eq}$. This is the regime
actually discussed by Onsager in his fundamental work
\cite{onsager1,onsager2}, in which he also discusses the celebrated
reciprocity relations involving the matrix $\mathcal{L}^* [{\textbf
a}^{eq}]$ of kinetic coefficients.

\subsection{Onsager-Machlup theory of thermal
fluctuations}\label{IV.2}

Onsager's  connection between the theory of irreversible processes
and the theory of spontaneous fluctuations was made by postulating
that ``the decay of a system from a given non-equilibrium state
produced by a spontaneous fluctuation $\delta {\textbf a }(t)\equiv
\textbf{a}(t)-{\textbf a }^{eq}$ obeys, \emph{on the average}, the
(empirical) law for the decay from the same state back to
equilibrium, when it has been produced by a constraint which is then
suddenly removed" \cite{onsagermachlup1}. This postulate, referred
to as the Onsager's regression hypothesis, along with the assumption
that the statistical properties of the fluctuations can be modeled
as an Ornstein-Uhlenbeck process, led Onsager, together with his
Ph.D. student S. Machlup, to develop their theory of time-dependent
fluctuations \cite{onsagermachlup1, onsagermachlup2}.

Such theory essentially consists of postulating that the dynamics of
$\delta {\textbf a }(t)$ is described by the stochastic version of
Eq. (\ref{linearreleq}), namely,
\begin{equation}
\frac{d\delta {\textbf a }(t)}{dt} = -\mathcal{L}^* [ {\textbf a
}^{eq}] \circ \mathcal{E}\left[{\textbf a }^{eq}\right]\circ \delta
{\textbf a}(t) + \textbf{f}(t), \label{oueq}
\end{equation}
where the stochastic vector $\textbf{f}(t)$ is assumed to be a
stationary, Gaussian, and purely random (or ``$\delta$-correlated")
noise, uncorrelated with the initial value $\delta {\textbf a }(0)$
of the fluctuations and with zero mean  and time-correlation
function given by $<\textbf{f}(t)\textbf{f}^{ \dagger}(t')>= \gamma
\delta (t-t')$. These assumptions, however, are equivalent to saying
that $\delta {\textbf a }(t)$ is an Ornstein-Uhlenbeck process, as
indicated in subsection \ref{III.1} (see Eq. (\ref{releq0001})).
Hence, automatically all the mathematical properties summarized in
section \ref{III} apply for the process $\delta {\textbf a }(t)$.
These include the time evolution of the conditional mean value
$\overline{\textbf {a}}^0(t)$ and the covariance $\sigma(t)$,
governed by the deterministic equations (\ref{releqmean}) and
(\ref{sigmadteq}), whose solutions in Eqs. (\ref{meansoloueq}) and
(\ref{solsigmadt}) interpolate between the given initial state
($\textbf{a}^0,\sigma^0$) and the final stationary state
($\textbf{a}^{ss},\sigma^{ss}$) (which in this case corresponds to
the equilibrium state ($\textbf{a}^{eq},\sigma^{eq}$)). These
properties, as well as the time-dependent correlation function
$C(t+\tau,t)$ in Eq. (\ref{cttp}), are thus written in terms of the
relaxation matrix $\mathcal{H^*}[{\textbf a}^{eq}]$ or, taking into
account Eq. (\ref{relaxmatrixh}), in terms of the kinetic matrix
$\mathcal{L}^* [{\textbf a}^{eq}]$ and the thermodynamic matrix
$\mathcal{E}\left[{\textbf a }^{\ eq}\right]$.

Summarized in the manner we have done in this and in the previous
subsection, Onsager's theory exhibits a given logical argumentation,
in which the foundation of the theory seems to be provided by a the
phenomenological laws of linear irreversible thermodynamics. In this
line of reasoning the fluctuations and their description enter as a
secondary topic, whose description needs just a few auxiliary
concepts, including that of an Ornstein-Uhlenbeck process. For our
purpose of generalizing Onsager's theory, however, an alternative
argumentation is not only conceptually more economical but also
easier to extend or generalize.

This alternative argumentation derives from only two fundamental
postulates. The first is the assumption that the thermal
fluctuations $\delta {\textbf a }(t)\equiv \textbf{a}(t)-{\textbf a
}^{eq}$ around the equilibrium state constitute an
Ornstein-Uhlenbeck stochastic process. The second postulate is that
the first two moments, $\textbf{a}^{ss}$ and $\sigma^{ss}$, of the
stationary distribution $W_1[{\textbf a }]$ of this
Ornstein-Uhlenbeck process, coincide with the first two moments
$\textbf{a}^{eq}$ and $\sigma^{eq}$ of the exact equilibrium
probability distribution function \cite{landaulifshitz}
\begin{equation}
W^{eq}[{\textbf a }]=e^{(S[{\textbf a }]-S[{\textbf a }^{eq}])/k_B},
\label{bpe}
\end{equation}
in which $S=S[{\textbf a }]$ is the entropy of the closed system.
From these two postulates all the results of Onsager's theory
follow.

To see this let us first recall \cite{callen,greenecallen} that from
the exact Boltzmann-Planck equilibrium distribution in Eq.
(\ref{bpe}) one can derive the value of all its moments, given the
fundamental thermodynamic relation $S=S[{\textbf a }]$. In
particular one can determine $\textbf{a}^{eq}$ and $\sigma^{eq}$.
Thus, since ${\textbf a }^{eq}$ corresponds to a maximum of the
entropy, it is determined by the well known and widely used first
equilibrium condition,
\begin{equation}
\label{firsteqcond} {\textbf F }[  {\textbf a}^{eq}] ={\textbf F
}^{eq},
\end{equation}
whereas $\sigma^{eq}$ is determined by the less well known but
equally important second equilibrium condition,
\begin{equation}\label{exsigma0}
\sigma^{eq}\circ {\mathcal E}[{\textbf a }^{eq}] = I,
\end{equation}
where $I$ is the $C$x$C$ identity matrix and with the matrix
${\mathcal E}[{\textbf a }]$ defined in Eq. (\ref{matrixE}).

The exact condition for the covariance in Eq. (\ref{exsigma0}) can
now be used in the mathematical result for the structure of the
relaxation matrix $\mathcal{H}$ in Eq. (\ref{matrixh}), which then
becomes identical to the expression for $\mathcal{H^*}$ in Eq.
(\ref{relaxmatrixh}). In this manner we identify the matrix
$\mathcal{L}$ of  Eq. (\ref{matrixh}) with the matrix $\mathcal{L}^*
[ {\textbf a }^{eq}]$ of Onsager's kinetic coefficients, and
conclude that in the present argumentation, the structure of the
linear relaxation equations in Eq. (\ref{linearreleq}) can be viewed
as a consequence of the two fundamental postulates above, and not as
a consequence of the phenomenological laws of linear irreversible
thermodynamics. Instead, we can actually invert the conventional
argument, and claim that the universal phenomenological validity of
linear irreversible thermodynamics provides the empirical validation
of the suitability of the Ornstein-Uhlenbeck model to describe the
thermal fluctuations near the equilibrium state. An important
challenge is now to explore possible manners to generalize Onsager's
approach to regimes not included in its original formulation. Our
proposal to do that is explained in the following section.

\section{Non-equilibrium extension of Onsager's theory}\label{V}

In contrast with the discussion of the previous section, in which we
had in mind the simpler scenario illustrated in Fig. 1(a), let us
now consider the schematic scenario illustrated by Fig. 1(b), in
which we have multiple stationary states corresponding to multiple
local entropy maxima, and hence, to multiple \emph{meta-stable}
thermodynamic equilibrium states. We first consider the application
of Onsager's theory to permanently meta-stable equilibrium states,
then to a (``quasi-static") sequence of meta-stable states in each
of which the system has sufficient time to equilibrate, and finally
to a sequence of meta-stable states in which the system does not
have sufficient time to equilibrate because the corresponding
relaxation times are longer than the time spent by the system  in
the basin of attraction of the instantaneously meta-stable states in
the sequence.

\subsection{Static entropy landscape and permanent meta-stable
states}\label{V.1}

Let us assume that the entropy landscape \emph{remains static} and
that, if the system was prepared in an initial state ${\overline
{\textbf a }}(0)={\textbf a }^0$ with $\sigma(0)=0$, and with
${\textbf a }^0$ contained in the basin of attraction ${\mathcal
T}({\textbf a }^{ss})$ of the stationary state ${\textbf a }^{ss}$,
the mean trajectory ${\overline{\textbf a}}^0(t)$ will be confined
to that basin and will relax eventually to the state ${\textbf a
}^{ss}$, in which it will remain indefinitely. We then postulate
that under these conditions the laws that govern the time evolution
of the system in this relaxation process are totally
indistinguishable from those that govern the relaxation towards the
most stable equilibrium state ${\textbf a }^{eq}$ when the system is
prepared in the basin of attraction ${\mathcal T}({\textbf a
}^{eq})$ of this most stable state. Since these laws are contained
in Onsager's theory, then our postulate is equivalent to the
assumption that all what we said in the previous section in the
context of the state ${\textbf a }^{eq}$ extends over to all the
other (meta-stable) equilibrium states ${\textbf a }^{ss}$ without
change, except for the replacement of the label ``\emph{eq}" by the
label ``\emph{ss}".

For example, the conditional probability density $P({\textbf a
}^0\mid{\textbf a },t)$ that the system is in the macroscopic state
${\textbf a }$ at time $t$, given that it was prepared in the
initial state ${\textbf a }^0\in{\mathcal T}({\textbf a }^{ss})$ at
time $t=0$ is given by the Gaussian distribution

\begin{equation}
P({\textbf a }^0\mid{\textbf a },t)= [(2\pi)^C \det{\sigma(t)}]
e^{-[({\textbf a }-{\overline{\textbf a}}^0(t))^{\dagger}\circ
\sigma ^{-1}(t)\circ ({\textbf a }-{\overline {\textbf a
}}^0(t)]/2}, \label{qsdistrfunct0}
\end{equation}
where the mean value ${\overline{\textbf a}}^0(t)$ satisfies Eq.
(\ref{linearreleq}), which we re-write as

\begin{equation}
\frac{d{\overline{\textbf a}}^0(t)}{dt} = -\mathcal{L}^* [{\textbf
a}^{ss}] \circ \mathcal{E}\left[ {\textbf a }^{ss}\right]\circ [
{\overline{\textbf a}}^0(t)-{\textbf a }^{ss}],
\label{sslinearreleq}
\end{equation}
whose solution is given by

\begin{equation}
\overline {\textbf a }^0(t) = {\textbf a }^{ss} + e^{- \mathcal{L}^*
[{\textbf a}^{ss}] \circ \mathcal{E}\left[{\textbf a}^{ss}\right]
t}\circ [{\textbf a }^{0}-{\textbf a }^{ss}], \label{solavss}
\end{equation}
with $\mathcal{E}\left[{\textbf a }^{ss}\right]$ defined as

\begin{equation}
\mathcal{E}_{ij}[{\textbf a }^{ss}] \equiv -\left( \frac{\partial
F_i[{\textbf a }]}{\partial a_j} \right)_{{\textbf a }={\textbf a
}^{ss}}=-\frac{1}{k_B}\left( \frac{\partial^2 S[{\textbf a
}]}{\partial a_i\partial a_j} \right)_{{\textbf a }={\textbf a
}^{ss}}. \label{ssmatrixE}
\end{equation}

Similarly, the covariance $\sigma(t)$ satisfies Eq.
(\ref{sigmadteq}) which, using Eqs. (\ref{fdtheorem0}) and
(\ref{relaxmatrixh}), can be written as

\begin{equation}
\frac{d\sigma(t)}{dt} = -\mathcal{L}^* [{\textbf a}^{ss}] \circ
\mathcal{E}\left[{\textbf a}^{ss}\right]\circ [\sigma
(t)-\sigma^{ss}] -k_B [\sigma (t)-\sigma^{ss}] \circ
\mathcal{E}\left[{\textbf a}^{ss}\right] \circ\mathcal{L}^{*\dagger
} [{\textbf a}^{ss}] .\label{sigmadtss}
\end{equation}
The corresponding solution, according to Eq. (\ref{solsigmadt}), is

\begin{equation}
\sigma(t)= \sigma^{ss} + e^{-\mathcal{L}^* [{\textbf a}^{ss}] \circ
\mathcal{E}\left[{\textbf a}^{ss}\right]t}\circ
[\sigma^{0}-\sigma^{ss}]\circ e^{(-\mathcal{L}^* [{\textbf a}^{ss}]
\circ \mathcal{E}\left[{\textbf a}^{ss}\right])^{\dagger }t} ,
\label{solsigmadtss}
\end{equation}
which interpolates the covariance $\sigma(t)$ between its initial
value $\sigma(0)=\sigma^0$ and its stationary value $\sigma^{ss}$
given, according to Eq. (\ref{exsigma0}), by the result

\begin{equation}\label{exsigma1}
 \sigma^{ss}= \left({\mathcal E}[{\textbf a }^{ss}]\right)^{-1}.
\end{equation}

Finally, according to Eqs. (\ref{dcttp}) and (\ref{cttp}), with the
matrix $\mathcal{H^*}$ given by Eq. (\ref{relaxmatrixh}), the
relaxation equation of the two-time correlation matrix
$C(t+\tau,t)\equiv <\delta\textbf{a}(t+\tau)
\delta\textbf{a}^\dagger(t) >$  is
\begin{equation}
\frac{dC(t+\tau,t)}{d\tau}=-\mathcal{L}^* [{\textbf a}^{ss}] \circ
\mathcal{E}\left[{\textbf a}^{ss}\right]\circ C(t+\tau,t),
\label{dcttpp}
\end{equation}
with initial condition $C(t+0,t)= \sigma(t)$, whose solution is
\begin{equation}
C(t+\tau,t)=e^{-\mathcal{L}^* [{\textbf a}^{ss}] \circ
\mathcal{E}\left[{\textbf a}^{ss}\right]\tau}\circ\sigma(t) \ \ \ \
\ \ \  (\textrm{for}\ \tau\ge0).  \label{cttpp}
\end{equation}

In this manner, if we knew the full thermodynamic landscape
described by the fundamental thermodynamic relation $S=S[{\textbf a
}]$, as well as the state dependence of the matrix $\mathcal{L}^*
[{\textbf a}]$ of Onsager kinetic coefficients, the results
summarized in this subsection would constitute the full and exact
solution of the Onsager-Machlup model of the irreversible relaxation
of a closed system towards any stationary state ${\textbf a }^{ss}$,
regardless if this is the most stable or any of the meta-stable
equilibrium states of the isolated system, but provided that the
system was prepared in an initial state ${\textbf a }^{0}$ that lies
inside the basin of attraction of this particular stationary state,
i.e., that ${\textbf a }^{0} \in {\mathcal T}({\textbf a }^{ss})$.

\subsection{Quasi-static relaxation to the most stable equilibrium
state ${\textbf a}^{eq}$}\label{V.2}

In the previous discussion we assumed that the entropy landscape in
Fig. 1(b) remains constant in time so that if the system was
prepared in the basin of attraction of a given stationary state, it
will be trapped in this basin, eventually relaxing to the
corresponding stationary state ${\textbf a}^{ss}$, to remain
indefinitely in that (stable or meta-stable) state. In reality, it
is impossible that such situation can be sustained indefinitely,
since there are many possible mechanisms that allow the trajectory
${\overline{\textbf a}}^0(t)$ to explore states outside the basin of
attraction in which the system was initially prepared. If such
excursion occurs to the basin of attraction of a stationary state
with a still larger local entropy maximum, the second law of
thermodynamics dictates that the system will proceed to the
stationary state of the new basin of attraction. If this process
occurs repeatedly, the system will eventually reach the most stable
equilibrium state ${\textbf a}^{eq}$, thus restoring to some extent
the irreversible process described in the simpler conditions
illustrated in Fig. 1(a), but with a probably much slower dynamics
because the transitions from basin to basin may involve activation
barriers that must await for adequate spontaneous fluctuations.

The irreversible process thus envisioned may in some sense be
similar to the concept of \textit{quasi-static process} in classical
thermodynamics. Thus, consider a system prepared in an initial
equilibrium state ${\textbf a }_I={\overline {\textbf a }}(0)$ that
evolves towards a final equilibrium state ${\textbf a }_{F}$ through
a sequence of intermediate equilibrium states $ {\textbf a
}_\alpha\equiv {\overline{\textbf a}}^0(t_\alpha)$ at time
$t_\alpha$, with $\alpha = 1,2,...$, aided by the sequential change
of the constraints imposed on the system, from those that
equilibrate the initial state ${\textbf a }_I$, to those that
equilibrate the state ${\textbf a}_1\equiv {\overline {\textbf a
}}^0(t_1)$, to those that equilibrate the state ${\textbf a}_2\equiv
{\overline{\textbf a}}^0(t_2)$,..., etc. The time evolution of the
covariance $\sigma(t)$ along this process is then described by the
interpolating expression in Eq. (\ref{solsigmadtss}) which, for the
time $t$ in the interval $t_{\alpha}\le t \le t_{\alpha +1}$, reads

\begin{eqnarray}
\sigma(t)= \mathcal{E}^{-1}_{\alpha +1} + e^{(-\mathcal{L}^*
_{\alpha +1} \circ \mathcal{E}_{\alpha +1})(t-t_{\alpha})}\circ
\left[ \sigma(t_{\alpha })-\mathcal{E}^{-1}_{\alpha +1} \right]\circ
e^{(-\mathcal{L}^* _{\alpha +1} \circ \mathcal{E}_{\alpha
+1})^{\dagger }(t-t_{\alpha})} \nonumber \\ \ \ \ (t_{\alpha}\le t
\le t_{\alpha +1}) \label{solsigmadtquasistatic}
\end{eqnarray}
where $\mathcal{E}_{\alpha +1}\equiv
\mathcal{E}\left[{\overline{\textbf a}}^0(t_{\alpha +1})\right]$ and
$\mathcal{L}^{*}_{\alpha +1}\equiv
\mathcal{L}^{*}\left[{\overline{\textbf a}}^0(t_{\alpha
+1})\right]$.

The main idealized property of a quasi-static process is, of course,
that in going from any intermediate equilibrium state ${\textbf
a}_\alpha$ to the next equilibrium state ${\textbf a}_{\alpha+1}$
the intervals $\mid t_{\alpha +1}-t_{\alpha}\mid$ are sufficiently
large to allow the system to equilibrate to the new state. This
requires these intervals to be much larger than any of the
relaxation times of the system, i.e., than any of the inverse
eigenvalues of the relaxation matrix $(\mathcal{L}^* _{\alpha +1}
\circ \mathcal{E}_{\alpha +1})$. The immediate consequence of this
assumption is that at the end $t_{\alpha+1}$ of that interval, the
covariance matrix $\sigma(t)$ has attained its equilibrium value
$\sigma(t_{\alpha +1})= \mathcal{E}^{-1}_{\alpha+1}$. Of course, if
the system relaxes infinitely fast, we can take these intervals to
be arbitrarily small, and then the covariance matrix attains its
``local equilibrium" value,

\begin{equation}
\sigma(t)= \left( \mathcal{E}\left[{\overline{\textbf
a}}^0(t)\right] \right)^{-1} \label{equilcondsigmap}
\end{equation}
at all times. A quasi-static process is thus characterized by the
assumption that this relation, referred to as the local equilibrium
approximation, is instantly valid at all times. Thus, for example,
the two-time correlation  matrix $C(t+\tau,t)$ would be given within
this approximation and according to Eq. (\ref{cttpp}), by
\begin{equation}
C(t+\tau,t)=e^{-\mathcal{L}^* [{\overline{\textbf a}}^0(t)] \circ
\mathcal{E}\left[{\overline{\textbf a}}^0(t)\right]\tau}\circ \left(
\mathcal{E}\left[{\overline{\textbf a}}^0(t)\right] \right)^{-1} \ \
\ \ \ \ \ (\textrm{for}\ \tau\ge0).  \label{cttppqs}
\end{equation}

In summary, a quasi-static process is characterized by an
instantaneously equilibrated probability distribution function,
i.e., by Eq. (\ref{qsdistrfunct0}) with $\sigma(t)$ given by its
local equilibrium value above, so that
\begin{equation}
P({\textbf a }^0\mid{\textbf a },t)= \left[(2\pi)^C /
\det{\mathcal{E}\left[{\overline{\textbf a}}^0(t)\right]}\right]
e^{-[({\textbf a }-{\overline{\textbf a}}^0(t))^{\dagger}\circ
\mathcal{E}\left[{\overline{\textbf a}}^0(t)\right]\circ ({\textbf a
}-{\overline{\textbf a}}^0(t))]/2}. \label{qsdistrfunct1}
\end{equation}
Unfortunately, real systems relax through real irreversible
processes that have little to do with this idealized concept.
Instead, the relaxation times assumed negligible in this limit are
surely finite and, under many circumstances, considerably long.
Furthermore, for experimental or other reasons, one may want to
consider intervals $\mid t_{\alpha +1}-t_{\alpha}\mid$ of much
shorter duration than the system's relaxation times. Hence, we have
in general that $\sigma(t)\ne \left( \mathcal{E}\left[{\overline
{\textbf a}}(t)\right] \right)^{-1}$, and an important fundamental
task is to describe and quantify the deviations from the idealized
limiting behavior involved in a quasi-static process. The simplest
approach to achieve this is also based in Eq.
(\ref{solsigmadtquasistatic}) above, as explained below.

\subsection{Out-of-equilibrium relaxation}\label{V.3}

Let us now go back to our closed system, which was prepared in an
initial state ${\textbf a }^0$ that is not the global equilibrium
state ${\textbf a }^{eq}$, but which relaxes irreversibly towards
this state of maximum global entropy. The statistical description of
the macroscopic state of this system is provided, as indicated
before, by the conditional probability density $P({\textbf a
}^0\mid{\textbf a },t)$ that the system is in the macroscopic state
${\textbf a }$ at time $t$, given that it was in the initial state
${\textbf a }^0$ at time $t=0$. Assuming that the thermal
fluctuations $\delta {\textbf a }(t)\equiv \textbf{a}(t)-{\overline
{\textbf a }}^0(t)$ can be modeled locally in time as an
Ornstein-Uhlenbeck process, the Gaussian property of this stochastic
process approximates $P({\textbf a }^0\mid{\textbf a },t)$ by the
Gaussian distribution $ P({\textbf a }^0\mid{\textbf a },t)=
[(2\pi)^C \det{\sigma(t)}] e^{-[({\textbf a }-{\overline {\textbf a
}}^0(t))^{\dagger}\circ \sigma ^{-1}(t)\circ ({\textbf a
}-{\overline {\textbf a }}^0(t)]/2}. $ Thus, the time evolution of
the macroscopic state is determined, within this simplified model,
by the time evolution of the mean value ${\overline{\textbf
a}}^0(t)$ and of the covariance $\sigma(t)$.

The mean value ${\overline{\textbf a}}^0(t)$ is supposed to solve an
equation of the general form in Eq. (\ref{releq0repet}), which can
also be written as $ {\overline{\textbf a}}^0(t+\Delta t)=
{\overline{\textbf a}}^0(t) +  \mathcal{R} [{\overline {\textbf a
}}(t)] \Delta t + \mathcal{O}(\Delta t)^2$. Thus, we may represent
the solution of this equation by a discrete sequence $ {\textbf a
}_{\alpha}\equiv {\overline{\textbf a}}^0(t_{\alpha})$ of the mean
value at the times $t_0(=0), t_1, ..., t_\alpha, ...,t_\infty
(=\infty)$, generated by the recurrence relation
\begin{equation}
{\textbf a }_{\alpha+1}= {\textbf a }_{\alpha} +  \mathcal{R}
[{\textbf a }_{\alpha}]\Delta t_\alpha , \label{releq0030}
\end{equation}
with the time intervals $\Delta t_\alpha\equiv t_{\alpha+1}
-t_\alpha$ chosen short enough for this linear approximation to be
valid.

At this point, we assume that the states ${\textbf a }_{\alpha}$ of
this sequence correspond to instantaneous local entropy maxima, and
hence, that within the interval $(t_{\alpha+1} -t_\alpha)$ the
system remains in the basin of attraction of the stationary state
${\textbf a }_{\alpha+1}$. We also assume that we can approximate
the corresponding fluctuations $\delta {\textbf a }(t)\equiv
\textbf{a}(t)-{\textbf a }_{\alpha+1}$ by an Ornstein-Uhlenbeck
process, so that in this time interval the time-evolution of the
covariance matrix $\sigma(t)$ is described precisely by Eq.
(\ref{solsigmadtquasistatic}). This is the expression employed in
the previous section to describe quasi-static processes. The main
difference is that now we are interested in the opposite limit, in
which the time intervals $\Delta t_\alpha\equiv t_{\alpha+1}
-t_\alpha$ are very short. Thus, we only need to consider Eq.
(\ref{solsigmadtquasistatic}) in its linear approximation in $(t
-t_\alpha)$. The resulting expression leads, with $t= t_{\alpha+1}$,
to the following recurrence relation for the sequence
$\sigma_\alpha\equiv \sigma (t_\alpha)$

\begin{equation}
\sigma_{\alpha+1} = \sigma_{\alpha} -\left[\mathcal{L}^* _{\alpha
+1} \circ \mathcal{E}_{\alpha +1}\circ \sigma_\alpha + \sigma_\alpha
\circ \mathcal{E}_{\alpha +1}\circ \mathcal{L}^{* \dagger} _{\alpha
+1} \right](\Delta t_{\alpha})+ \left[\mathcal{L}^* _{\alpha +1} +
\mathcal{L}^{* \dagger} _{\alpha +1} \right](\Delta t_{\alpha}).
\label{releq003}
\end{equation}
where, as before, $\mathcal{E}_{\alpha }\equiv
\mathcal{E}\left[{\overline{\textbf a}}^0(t_{\alpha })\right]$ and
$\mathcal{L}^{*}_{\alpha }\equiv
\mathcal{L}^{*}\left[{\overline{\textbf a}}^0(t_{\alpha})\right]$.
This, however, is nothing but the discrete version of the following
differential equation for $\sigma(t)$

\begin{equation}
\frac{d\sigma(t)}{dt} = -\mathcal{L}^* [{\overline {\textbf a
}}^0(t)] \circ \mathcal{E}\left[{\overline {\textbf a
}}(t)\right]\circ \sigma (t) - \sigma (t) \circ
\mathcal{E}\left[{\overline{\textbf a}}^0(t)\right]
\circ\mathcal{L}^{*\dagger } [{\overline{\textbf a}}^0(t)] + \left(
\mathcal{L}^* [{\overline{\textbf a}}^0(t)] + \mathcal{L}^{* \dagger
} [{\overline{\textbf a}}^0(t)] \right).\label{sigmadtirrev}
\end{equation}

In contrast with Eq. (\ref{sigmadtss}), which has the simple
analytic solution of Eq. (\ref{solsigmadtss}), no analytic solution
exists for this equation, and hence, the numerical calculation of
$\sigma(t)$ may be based on the recursion relation in Eq.
(\ref{releq003}). The difference between the resulting time
dependent covariance $\sigma(t)$ and its local equilibrium value
$(\mathcal{E}\left[{\overline{\textbf a}}^0(t)\right])^{-1}$
measures the departure of the actual irreversible process from the
idealized quasi-static process discussed in the previous subsection.

\subsection{Locally stationary two-time correlation function}\label{V.4}

Another important measurable property is the two-time correlation
matrix $C(t',t)\equiv \overline{\delta {\textbf a }(t')\delta
{\textbf a }^{ \dagger}(t)}$, where now  $\delta {\textbf a }(t)
\equiv \textbf{a}(t)-{\overline{\textbf a}}^0(t)$ refers to the
fluctuations around the time-evolving mean value ${\overline
{\textbf a }}(t)$. Thus defined, the fluctuations $\delta {\textbf a
}(t)$ do not in general constitute a stationary stochastic process.
However, just like we did in the derivation of the recurrence
relation in Eq. (\ref{releq003}), here we also assume the
\emph{local stationarity} approximation, i.e., we assume that in the
interval $t_{\alpha}\le t \le t_{\alpha +1}$, the mean value
${\overline {\textbf a }}^0(t_\alpha)$ and the covariance
$\sigma(t_\alpha)$ can be considered approximately constant,
${\overline {\textbf a }}^0(t)\approx{\overline{\textbf
a}}^0(t_\alpha)$ and $\sigma(t)\approx \sigma(t_\alpha)$, so that
the thermal fluctuations $\delta {\textbf a }(t_\alpha+\tau) \equiv
{\textbf a }(t_\alpha+\tau)-{\overline{\textbf a}}^0(t_\alpha)$
around the momentarily stationary value ${\overline{\textbf
a}}^0(t_\alpha)$ can be described approximately as an
Ornstein-Uhlenbeck process.

Clearly, the physical notion behind this assumption is that both,
${\overline{\textbf a}}^0(t_\alpha)$ and $\sigma(t_\alpha)$, are
macroscopic variables that relax to their equilibrium values within
rather slow \emph{macroscopic relaxation times} described by the
time coordinate $t_\alpha$, whereas the thermal fluctuations $\delta
{\textbf a }(t_\alpha+\tau)$ reflect much more local and faster
microscopic events, whose correlations decay within
\emph{microscopic correlation times} described by the time $\tau$.
These faster events would be averaged out when observing only
${\overline{\textbf a }}^0(t_\alpha)$ and $\sigma(t_\alpha)$, and
will only ``renormalize" the slow decay of these properties. Their
direct measurement, however, is possible through the measurement of
the $\tau$-dependence of the correlation function $C(t_\alpha
+\tau,t_\alpha)\equiv \overline{\delta {\textbf a }(t_\alpha
+\tau)\delta {\textbf a }^{ \dagger}(t_\alpha)}$.

From Eq. (\ref{otravez}) we may write the stochastic equation for
the locally stationary fluctuations $\delta {\textbf a
}(t_\alpha+\tau)$ as
\begin{equation}
\frac{d\delta   {\textbf a }(t_\alpha +\tau)}{d\tau} =
\mathcal{H}(t_\alpha) \circ \delta {\textbf a}(t_\alpha +\tau) +
{\textbf f}(t_\alpha +\tau) \label{localsslinearreleq}
\end{equation}
with the relaxation matrix $\mathcal{H}(t_\alpha)$ given, according
to the general stationary condition in Eq. (\ref{matrixh}), by
\begin{equation}
\mathcal{H}(t_\alpha) = - \mathcal{L}(t_\alpha) \circ
\sigma^{-1}(t_\alpha) \label{hlocalss}
\end{equation}
Multiplying Eq. (\ref{localsslinearreleq}) on the right by the
transpose of $\delta {\textbf a }(t_\alpha+0)$, and averaging over
the realizations of ${\textbf f}(t_\alpha +\tau)$ and over the
``initial" values $\delta {\textbf a }(t_\alpha+0)$, we derive the
following relaxation equation for $C(t_\alpha +\tau,t_\alpha)\equiv
\overline{\delta {\textbf a }(t_\alpha +\tau)\delta {\textbf a }^{
\dagger}(t_\alpha)}$
\begin{equation}
\frac{dC(t_\alpha +\tau,t_\alpha)}{d\tau} =  - \mathcal{L}(t_\alpha)
\circ \sigma^{-1}(t_\alpha) \circ C(t_\alpha +\tau,t_\alpha)
\label{localssdcdt}
\end{equation}
whose solution is
\begin{equation}
C(t_\alpha +\tau,t_\alpha) = e^{ - \mathcal{L}(t_\alpha) \circ
\sigma^{-1}(t_\alpha) \tau} \circ \sigma(t_\alpha ).
\label{localsscdt}
\end{equation}

This expression for $C(t_\alpha +\tau,t_\alpha)$ should be compared
with the result in Eq. (\ref{cttpp}), which was derived from Eq.
(\ref{cttp}) with the relaxation matrix $\mathcal{H}$ given,
however, by Eq. (\ref{relaxmatrixh}), whose validity is restricted
to the linear regime of the stationary state ${\overline {\textbf a
}}(t_\alpha)$. Notice also that in the local equilibrium
approximation, $\sigma(t) =\left( \mathcal{E}\left[{\overline
{\textbf a}}(t)\right] \right)^{-1}$, the result in Eq.
(\ref{localsscdt}) coincides with the quasi-static expression in Eq.
(\ref{cttppqs}).

\subsection{Generalized Ornstein-Uhlenbeck processes}\label{V.5}

As discussed above, the $\tau$-dependence of the thermal
fluctuations $\delta {\textbf a }(t_\alpha+\tau)$ describes faster
microscopic events, whose correlations decay within
\emph{microscopic correlation times}. When viewed with this temporal
resolution, however, it is mandatory to revise the assumption that
the random term ${\textbf f}(t_\alpha +\tau)$ in Eq.
(\ref{localsslinearreleq}) can be approximated by a
$\delta$-correlated noise, an assumption that allowed the dynamics
of the thermal fluctuations to be described by an Ornstein-Uhlenbeck
process. Relaxing this fundamental assumption requires the
definition of a more general mathematical model of a stationary
stochastic process which is not required to be necessarily
Markovian.

Such mathematical model, which we will refer to as a
\emph{generalized} Ornstein-Uhlenbeck process, was discussed in Ref.
\cite{delrio} and allows us, in the present context, to describe
$\delta {\textbf a }(t_\alpha +\tau)$ in terms of the most general
linear stochastic differential equation with additive noise, namely,
\begin{equation}
\frac{d\delta  {\textbf a }(t_\alpha +\tau)}{d\tau}= -
\int_0^{\tau}d\tau' \mathcal{H}(t_\alpha; \tau-\tau')\circ \delta
{\textbf a }(t_\alpha +\tau')+{\textbf f }(t_\alpha +\tau).
\label{fluctuations00}
\end{equation}
In this model, the stochastic vector $\textbf{f}(t_\alpha +\tau)$ is
only assumed to be necessarily stationary but not necessarily
Gaussian nor white; it is still assumed uncorrelated with the
initial value $\delta {\textbf a }(t_\alpha +0)$ and to have zero
mean, $ {\overline {\textbf f }}(t_\alpha +\tau)=\textbf{0}$.

According to the theorem of stationarity \cite{delrio,faraday}, the
stationarity condition is mathematically equivalent to the condition
that Eq. (\ref{fluctuations00}) conforms to a very strict and rigid
format, namely,
\begin{equation}
\frac{d\delta  {\textbf a }(t_\alpha +\tau)}{d\tau}= -
\omega(t_\alpha ) \circ [ \sigma(t_\alpha)]^{-1} \circ \delta
{\textbf a }(t_\alpha +\tau) - \int_0^{\tau}d\tau' L(t_\alpha
;\tau-\tau')\circ [ \sigma(t_\alpha)]^{-1}\circ \delta {\textbf a
}(t_\alpha +\tau')+{\textbf f }(t_\alpha +\tau)
\label{fluctuations1}
\end{equation}
where $\omega(t_\alpha )$ is an antisymmetric matrix,
$\omega(t_\alpha )=-\omega^{ \dagger}(t_\alpha )$, the memory
function $L(t_\alpha ;\tau)$ satisfies the following
fluctuation-dissipation relation

\begin{equation}
L(t_\alpha ;\tau) = L^{\dagger}(t_\alpha ;-\tau)= <{{\textbf f
}(t_\alpha +\tau){\textbf f }^{\ \dagger}(t_\alpha +0)}>;
\label{fluct-disip}
\end{equation}
the matrix $\sigma(t_\alpha)\equiv  {\overline{\delta{\textbf
a}(t_\alpha+0)\delta{\textbf a }^{\dagger}(t_\alpha+0)}}$ is the
covariance of the probability distribution function of the initial
value $\delta {\textbf a }(t_\alpha+0)$ of the fluctuations. In
other words, the non-Markovian relaxation matrix $
\mathcal{H}(t_\alpha; \tau)$  of Eq. (\ref{fluctuations00}) is not
arbitrary, but \emph{must} have the rigid format leading to Eq.
(\ref{fluctuations1}), namely,
\begin{equation}
\mathcal{H}(t_\alpha; \tau)= - \left[2\delta(\tau)\omega (t_\alpha)
+ L(t_\alpha; \tau)\right]\circ [ \sigma(t_\alpha)]^{-1}.
\label{hdass}
\end{equation}

The generalized Ornstein-Uhlenbeck model defined by Eq.
(\ref{fluctuations1}) may be used to describe the fluctuations
around \emph{any} stationary state, including meta-stable and
absolutely stable thermodynamic equilibrium states. Under stationary
conditions the label $t_\alpha$ is, of course, unnecessary and Eq.
(\ref{fluctuations1}) may be recognized in a statistical mechanical
context \cite{berne} as the \emph{ generalized Langevin equation}
(GLE). In that context, however, the term GLE is associated with the
stochastic equation formally derived from a N-particle microscopic
(Newtonian or Brownian) dynamic description by means of projection
operator techniques to describe the time-dependent thermal
fluctuations of systems in thermodynamic equilibrium \cite{berne}.
Indeed, such an equation conforms exactly to the format described by
Eq. (\ref{fluctuations1}). It is important to insist, however, that
this format has a purely mathematical origin, imposed solely by the
stationarity condition, and is certainly NOT a consequence of the
formal possibility of deriving it from an underlying microscopic
level of description. In fact, it is this mathematical structure of
the GLE, and the ``selection rules" imposed by the symmetry
properties of the matrices $\omega(t_\alpha)$ and $L(t_\alpha;
\tau)$ (along with other selection rules imposed by other possible
symmetries \cite{delrio}), what allows a fruitful use of the rigid
format of this equation to describe complex dynamic phenomena in a
rather simple manner, with virtually complete independence of the
detailed N-particle microscopic dynamics underlying the
time-evolution of the fluctuations $\delta {\textbf a }(t)$.

Multiplying Eq. (\ref{fluctuations1}) on the right by $\delta
{\textbf a }^{ \dagger}(t_\alpha)$, and taking the corresponding
average, we have that the two-time correlation function $C(t_\alpha
+\tau,t_\alpha)\equiv \overline{\delta {\textbf a }(t_\alpha
+\tau)\delta {\textbf a }^{ \dagger}(t_\alpha)}$, that we shall now
denote as $C_{t_\alpha}(\tau)$, satisfies the following relaxation
equation
\begin{equation}
\frac{dC_{t_\alpha}(\tau)}{d\tau}= - \omega(t_\alpha ) \circ [
\sigma(t_\alpha)]^{-1} \circ C_{t_\alpha}(\tau) -
\int_0^{\tau}d\tau' L(t_\alpha ;\tau-\tau')\circ [
\sigma(t_\alpha)]^{-1}\circ C_{t_\alpha}(\tau') \label{dcttaudtau}
\end{equation}
whose solution may be written, in terms of the Laplace transforms
$\hat{C}_{t_\alpha}(z)$ and $\hat{L}({t_\alpha};z)$ of
$C_{t_\alpha}(\tau)$ and $L({t_\alpha};\tau)$, as
\begin{equation}
\hat{C}_{t_\alpha}(z)=\left\{ z\textbf{I} + [
\omega({t_\alpha}) + \hat{L}({t_\alpha};z) ]\circ [
\sigma(t_\alpha)]^{-1}\right\}^{-1}\circ \sigma(t_\alpha).
\label{solfluctuations2}
\end{equation}

Let us notice that the results in this subsection reduce to the
results of the previous subsection in the Markovian limit, defined
by the condition $\hat{L}({t_\alpha};z) \approx
\hat{L}({t_\alpha};z=0)\equiv L^0({t_\alpha})$ or, equivalently, by
the condition $ L({t_\alpha};\tau)\approx
2\delta(\tau)L^0({t_\alpha})$, with $ L^0({t_\alpha})\equiv
\int_0^\infty dt L({t_\alpha};\tau)$. Thus, it is not difficult to
see that Eqs. (\ref{fluctuations1}), (\ref{dcttaudtau}), and
(\ref{solfluctuations2}) above become, respectively, Eqs. (
\ref{localsslinearreleq}), (\ref{localssdcdt}), and
(\ref{localsscdt}) of the previous section, and that Eq.
(\ref{hdass}) for $\mathcal{H}(t_\alpha;\tau)$ corresponds, after
integrating over $\tau$, to Eq. (\ref{hlocalss}) for
$\mathcal{H}(t_\alpha)$ upon the identification of the matrix
$\mathcal{L}(t_\alpha)$ with
\begin{equation}
\mathcal{L}(t_\alpha)= \omega(t_\alpha) + L^0(t_\alpha).
\label{okcss}
\end{equation}

Finally, let us postulate a ``correspondence principle" that
guarantees that in the vicinity of the stable thermodynamic
equilibrium state $\overline {\textbf a }^{eq}$, the present
nonlinear theory reduces to Onsager's original linear theory. For
this, we assume that the matrices $\omega(t_\alpha)$ and $L(t_\alpha
;\tau)$ depend on the relaxation time $t_\alpha$ only through
$\overline{\textbf{a}}(t_\alpha)$, so that $\omega (t_\alpha)=
\omega[\overline{\textbf{a}}(t_\alpha)]$ and $L(t_\alpha;\tau)=
L[\tau;\overline{\textbf{a}}(t_\alpha)]$. Then, Eq. (\ref{okcss})
may be rewritten as
\begin{equation}
\mathcal{L}(t_\alpha)= \omega[\overline{\textbf{a}}(t_\alpha)] +
\int_0^{\infty} d\tau L[\tau;\overline{\textbf{a}}(t_\alpha)].
\label{okcssp}
\end{equation}
We then postulate that $\mathcal{L}(t_\alpha)$ must coincide, when
the system has fully relaxed to the equilibrium state
$\textbf{a}^{eq}$,  with the phenomenological matrix
$\mathcal{L}^*[\textbf{a}^{eq}]$ of Onsager's kinetic coefficients
involved in the linear laws of Eq. (\ref{linearreleq}), i.e., that

\begin{equation}
\mathcal{L}^*[\textbf{a}^{eq}]= \omega[\textbf{a}^{eq}] +
\int_0^{\infty} d\tau L[\tau;\overline{\textbf{a}}^{eq}].
\label{okcsspp}
\end{equation}

\section{General description of out-of-equilibrium
relaxation}\label{VI}

In this section we recapitulate the discussion of the previous
section as a final proposal for the general canonical description of
the nonlinear non-equilibrium relaxation of macroscopic systems.
This scheme is summarized by the time-evolution equations for the
mean value $\overline{\textbf{a}}(t)$, for the covariance $\sigma
(t)$, and for the time-dependent correlation function
$C_t(\tau)\equiv \overline{\delta {\textbf a }(t+\tau)\delta
{\textbf a }^{ \dagger}(t)}$.

Let us first mention that an essential piece of information that
must be provided externally to this canonical theory is the
fundamental thermodynamic relation $S=S[\textbf{a}]$. From this
relation one is supposed to determine the thermodynamic equations of
state, i.e., the functional dependence on \textbf{a} of the
intensive variables
\begin{equation} \label{intensive}
F_i[\textbf{a}]\equiv k_B^{-1}(\partial S[\textbf{a}]/
\partial a_i), \end{equation}
as well as the thermodynamic matrix $\mathcal{E}\left[ {\textbf a
}\right]$, defined as
\begin{equation}
\mathcal{E}_{ij}[{\textbf a }] \equiv -\left( \frac{\partial
F_i[{\textbf a }]}{\partial a_j} \right)=-\frac{1}{k_B}\left(
\frac{\partial^2 S[{\textbf a }]}{\partial a_i\partial a_j} \right).
\label{matrixE}
\end{equation}
A stable thermodynamic equilibrium state is determined by the
equilibrium condition for ${\textbf a}^{eq}$ in Eq.
(\ref{firsteqcond}), namely,
\begin{equation}
\label{firsteqcondp} {\textbf F }[  {\textbf a}^{eq}] ={\textbf F
}^{eq},
\end{equation}
and by the condition in Eq. (\ref{exsigma0}) that determines the
equilibrium value $\sigma^{eq}$ of the covariance, namely,
\begin{equation}\label{exsigma0p}
\sigma^{eq}\circ {\mathcal E}[{\textbf a }^{eq}] = I,
\end{equation}

In the neighborhood of a thermodynamic equilibrium state, the
generally nonlinear relaxation equation for the conditional mean
value ${\overline{\textbf a}}^0(t)$  can be linearized in the
difference $\Delta {\overline {\textbf a }}(t) \equiv
{\overline{\textbf a}}^0(t)- {\textbf a}^{eq}$, to read

\begin{equation}
\frac{d\Delta  {\overline{\textbf a}}(t)}{dt} =
\mathcal{H^*}[\textbf{a}^{eq}]\circ \Delta {\overline {\textbf
a}}(t), \label{linearreleqh}
\end{equation}
with the matrix $\mathcal{H^*}[\textbf{a}^{eq}]$ given by
\begin{equation}
\mathcal{H^*} [{\textbf a}^{eq}]= -\mathcal{L}^* [{\textbf a}^{eq}]
\circ \mathcal{E}\left[{\textbf a }^{\ eq}\right],
\label{relaxmatrixh}
\end{equation}
with $\mathcal{L}^* [{\textbf a}^{eq}]$ being the matrix of
Onsager's kinetic coefficients.

Outside this linear regime, however, the mean value
$\overline{\textbf{a}}(t)$ is governed by the phenomenological
transport equation for the corresponding macroscopic state
variables, written in general, according to Eq. (\ref{releq0repet}),
as

\begin{equation}
\frac{d{\overline{\textbf a}}(t)}{dt}= \mathcal{R}\left[{\overline
{\textbf a }}(t)   \right]. \label{releq0repetpp}
\end{equation}
The constitutive relation, which determines the generally nonlinear
dependence of the vector $\mathcal{R}\left[{\overline {\textbf a
}}(t) \right]$ on the state vector ${\overline {\textbf a }}(t)$, is
in general unknown and is regarded as another external input of the
present theory. In the vicinity of a thermodynamic equilibrium state
${\textbf a}^{eq}$, however, this non-linear dependence must reduce
to the linear laws of irreversible thermodynamics.

Assuming that the matrices $\mathcal{L}^* [{\textbf a}]$ and
$\mathcal{E}\left[ {\textbf a }\right]$ are defined at any
accessible state \textbf{a}, the time-evolution of the covariance
matrix $\sigma(t)$ is governed by Eq. (\ref{sigmadtirrev}), i.e.,
\begin{equation}
\frac{d\sigma(t)}{dt} = -\mathcal{L}^* [{\overline {\textbf a
}}^0(t)] \circ \mathcal{E}\left[{\overline {\textbf a
}}(t)\right]\circ \sigma (t) - \sigma (t) \circ
\mathcal{E}\left[{\overline{\textbf a}}^0(t)\right]
\circ\mathcal{L}^{*\dagger } [{\overline{\textbf a}}^0(t)] + \left(
\mathcal{L}^* [{\overline{\textbf a}}^0(t)] + \mathcal{L}^{* \dagger
} [{\overline{\textbf a}}^0(t)] \right),\label{sigmadtirrev2}
\end{equation}
where ${\overline{\textbf a}}^0(t)$ is the solution of the nonlinear
equation (\ref{releq0repetpp}). This equation for $\sigma(t)$ has no
explicit solution, and hence, must be solved simultaneously with Eq.
(\ref{releq0repetpp}) for ${\overline {\textbf a }}(t)$ using, for
example, the recursion relations in Eqs. (\ref{releq0030}) and
(\ref{releq003}).

The dynamics of the locally stationary fluctuations $\delta {\textbf
a }(t+\tau)= {\textbf a }(t+\tau)-\overline{{\textbf a }}^0(t)$
around the conditional mean value $\overline{{\textbf a }}^0(t)$ are
described by Eq. (\ref{fluctuations1}), which reads

\begin{equation}
\frac{\partial \delta {\textbf a }(t+\tau)}{\partial \tau}= -
\omega[\overline{\textbf{a}}(t)]\circ \sigma^{-1} (t)\circ \delta
{\textbf a }(t+\tau) -\int_0^\tau d\tau '
L[\tau-\tau';\overline{\textbf{a}}(t)] \circ \sigma^{-1}(t)\circ
\delta {\textbf a }(t+\tau') + \textbf{f}(t+\tau),
\label{fluctuations2}
\end{equation}
with $<\textbf{f}(t+\tau)>=0$ and
$<\textbf{f}(t+\tau)\textbf{f}^{\dagger}(t+\tau')>
=L[\tau-\tau';\overline{\textbf{a}}(t)]$. From this equation one
derives the time-evolution equation of the non-stationary time
correlation function $C_t(\tau)\equiv <\delta {\textbf a
}(t+\tau)\delta {\textbf a }^{ \dagger}(t)>$, which reads
\begin{equation}
\frac{\partial C_{t}(\tau)}{\partial \tau}= -
\omega[\overline{\textbf{a}}(t)]\circ \sigma^{-1} (t)\circ
C_{t}(\tau) -\int_0^\tau d\tau '
L[\tau-\tau';\overline{\textbf{a}}(t)] \circ \sigma^{-1}(t)\circ
C_{t}(\tau'),  \label{fluctuations3}
\end{equation}
whose initial condition is $C_{t}(\tau=0)=\sigma(t)$. This equation
describes the decay of the correlation function $C_{t}(\tau)$ with
the ``microscopic" correlation time $\tau$, after the system has
evolved during a ``macroscopic" relaxation time $t$ from an initial
state described by $ \textbf{a}^0 \equiv \overline{\textbf{a}}(t=0)$
and $\sigma^0\equiv \sigma(t=0)$, to the ``current" state described
by $\overline{\textbf{a}}(t)$ and $\sigma(t)$.

Finally, we postulate a connection between the phenomenological
matrix $\mathcal{L}^*[\textbf{a}^{eq}]$ of Onsager's kinetic
coefficients involved in the linear laws of Eqs.
(\ref{linearreleqh}) and (\ref{relaxmatrixh}), and the matrices
$\omega[\textbf{a}]$ and $L[\tau;\textbf{a}]$ that describe,
according to Eq. (\ref{fluctuations2}), the dynamics of the
fluctuations $\delta {\textbf a }(t+\tau)$. This connection,
established in Eq. (\ref{okcsspp}) above, reads in general
\begin{equation}
\mathcal{L}^*[\textbf{a}]= \omega[\textbf{a}] + \int_0^{\infty}
d\tau L[\tau;\textbf{a}], \label{okcssppp}
\end{equation}
and constitutes a correspondence principle which ensures that the
theory just summarized contains the conventional Onsager's theory as
a particular case in the vicinity of a thermodynamic equilibrium
state.

In this manner, for given initial conditions
$\overline{\textbf{a}}(t=0)= {\textbf{a}}^0$, $\sigma
(t=0)=\sigma^0$, and $C_t(\tau=0)=\sigma(t)$, Eqs.
(\ref{releq0repetpp}), (\ref{sigmadtirrev2}), and
(\ref{fluctuations3}), together with the relationship in Eq.
(\ref{okcssppp}), would constitute a closed system of equations if
two fundamental pieces of information were available. The first is
the fundamental thermodynamic relation $S=S[\textbf{a}]$, from which
the state-dependence of ${\textbf F }[ {\textbf a}]=(\partial
S[\textbf{a}]/\partial \textbf{a})$ and $\mathcal{E}\left[ {\textbf
a }\right]=-(\partial \textbf{F}[\textbf{a}]/\partial
\textbf{a})/k_B$ could be derived. The second refers to the
conservative and dissipative kinetic matrices, $\omega[{\overline
{\textbf a }}(t)]$ and $L[\tau;{\overline {\textbf a }}(t)]$,
entering in Eq. (\ref{fluctuations3}). These two fundamental pieces
of information must be provided externally to the general format
above, and in many cases, their investigation constitutes a relevant
problem by itself. However, for a given specific physical context,
the format just described may guide us in the construction of the
specific models and approximations that best suit the description of
a particular relaxation phenomenon. For example, the application of
the generalized Onsager's theory just summarized, to the specific
context of the dynamics of colloidal dispersions, was spelled out in
Ref. \cite{eom2}. In order to illustrate the concrete use of the
general theory in a concrete physical condition, in what follows we
present a brief review of some aspects of such application.

\section{Application to colloid dynamics}\label{VII}

In this section we apply the general concepts above, to the specific
problem of the diffusive relaxation of the local concentration of
particles in a colloidal dispersion without hydrodynamic
interactions. Thus, let us consider a dispersion of $N$ colloidal
particles of mass $M$ in a volume $V$ which, in the absence of
external fields, has a uniform bulk number concentration $n_B =
N/V$. In the presence of a conservative static external field that
exerts a force $\textbf{F}^{ext} (\textbf{r}) = - \nabla \Psi
(\textbf{r}) $ on one particle located at position $\textbf{r}$, the
mean local concentration profile of colloidal particles,
$\overline{n} (\textbf{r},t)$, will evolve in time from some initial
condition $\overline{n }(\textbf{r},t=0)=n^0 (\textbf{r})$ towards
its stable thermodynamic equilibrium value $n^{eq} (\textbf{r})$.
The initial profile $n^0 (\textbf{r})$ is, of course, arbitrary,
whereas the final equilibrium profile $n^{eq} (\textbf{r})$ is
univocally dictated by the external  and internal constraints on the
system according to the second law of thermodynamics. In practice,
the external constraints are represented by the potential $\Psi
(\textbf{r}) $ of the external forces, while the internal
constraints originate in the intermolecular interactions,
represented by a pair potential $u(\textbf{r}, \textbf{r}')$.

To simplify the correspondence with the general theory of the
previous section, let us imagine that we divide the volume $V$
occupied by the colloidal dispersion in $C$ cells of equal volume
$\Delta V$ fixed in space. We then describe the macroscopic state
${\textbf a }(t)$ of this system in terms of the variables
$a_i(t)=n_i(t)\equiv N_i(t)/\Delta V$, where $N_i(t)$ is the number
of colloidal particles in cell $i$ at time $t$. We shall employ the
results of the previous section, however, in the understanding that
the continuum limit, $C \to \infty$ and $\Delta V \to 0$, has been
taken. In this limit, the label $i$ $(=1,2,...,C)$ of the component
$a_i$ is changed to the label $\textbf{r} \in V$ denoting the
spatial position of the center of the cell, and the component
$a_i(t)$ becomes the function $n(\textbf{r} ,t)$, which is the local
concentration profile of colloidal particles at time $t$.

Let us first discuss the application of the general equilibrium
conditions in Eqs. (\ref{firsteqcondp}) and (\ref{exsigma0p}) that
determines, in the equilibrium state, the mean value $n^{eq}
(\textbf{r})$ and the covariance matrix
$\sigma^{eq}(\textbf{r},\textbf{r}')$, and then we identify the
kinetic information by means of a phenomenological derivation of the
generalized diffusion equation.

\subsection{Fundamental thermodynamic relation and equilibrium conditions}\label{VII.1}

The fundamental thermodynamic relation of the present system
expresses the functional dependence of the entropy $S$ on the local
concentration profile $n(\textbf{r})$, a dependence represented by
$S=S[n]$. The intensive variable conjugate of $n(\textbf{r})$ is
$-\beta\mu [{\bf r};n]$, i.e., it is the negative of the local
electrochemical potential $\mu [{\bf r};n(t)] $  at position ${\bf
r}$ in units of the thermal energy $k_BT=\beta^{-1}$. It is an
ordinary function of ${\bf r}$ and a \emph{functional} of the
concentration profile $n(\textbf{r})$, written in general as
\cite{evans}

\begin{eqnarray}\label{1}
\beta\mu [{\bf r};n] & & =
\beta\mu^{in} [{\bf r};n] +  \beta \Psi({\bf r}) \\
&& \equiv  \beta\mu^{*}(\beta) + \ln n({\bf r}) -c[{\bf r};n] +
\beta \Psi({\bf r}).\nonumber
\end{eqnarray}
The first two terms of this definition of $\mu^{in} [{\bf r};n]$,
$(\beta\mu^{*}(\beta) + \ln n({\bf r}))$, are the ideal gas
contribution to the chemical potential, whereas  the term $-c[{\bf
r};n]$ contains the deviations from ideal behavior due to
interparticle interactions.

The equilibrium condition in Eq. (\ref{firsteqcondp}), which
determines $n^{eq}({\bf r})$, may then be written as
\begin{equation}
\beta\mu [{\bf r};n^{eq}] = \beta\mu^{*}(\beta) + \ln n^{eq}({\bf
r}) -c[{\bf r};n^{eq}] + \beta \Psi({\bf r}) = \beta\mu ^R \label{2}
\end{equation}
where the constant $\mu ^R $ is the uniform value of the
electrochemical potential. This would be a closed equation for
$n^{eq}({\bf r})$ if we knew the functional dependence of $c[{\bf
r};n]$ on $n({\bf r})$. The simplest example of a proposed
approximate functional dependence is the linear functional $c[{\bf
r};n]=-\int d \textbf{r}' \beta u(\textbf{r}, \textbf{r}')n
(\textbf{r}')$, referred to as the Debye-H\"uckel approximation.

The thermodynamic matrix $\mathcal{E}\left[ {\textbf a }\right]$
defined in Eq. (\ref{matrixE}) can be written in general, using Eq.
(\ref{1}),  as

\begin{equation}
\mathcal{E}[{\bf r},{\bf r}';n]\equiv \left[\frac {\delta \beta\mu
[{\bf r};n]}{\delta n({\bf r}')}\right] = \delta({\bf r}-{\bf r}')/
n({\bf r}) -c^{(2)}[{\bf r},{\bf r}';n], \label{stabmatrix}
\end{equation}
with $c^{(2)}[{\bf r},{\bf r}';n]\equiv (\delta c[{\bf r};n]/\delta
n({\bf r}'))$ being the \emph{functional }derivative of $c[{\bf
r};n]$ with respect to $n({\bf r}')$, referred to as the
\emph{direct correlation function}. On the other hand, the
covariance matrix $\sigma({\bf r},{\bf r}')= \overline{\delta n({\bf
r},0)\delta n({\bf r}',0)}$ can be written in terms of the
\emph{total correlation function} $h^{(2)}({\bf r},{\bf r}')$ as

\begin{equation}
\sigma({\bf r},{\bf r}') = n({\bf r}) \delta({\bf r}-{\bf r}') +
n({\bf r})n({\bf r}')h^{(2)}({\bf r},{\bf r}'). \label{covarmatrix}
\end{equation}
The matrices $\mathcal{E}[{\bf r},{\bf r}';n]$ and $\sigma({\bf
r},{\bf r}')$ are not in general related to each other. When
evaluated at the equilibrium local concentration profile
$n^{eq}({\bf r})$, however, they are related by the second
equilibrium condition  in Eq. (\ref{exsigma0p}), which in the
present context reads

\begin{equation}
\int d\textbf{r}'\sigma^{eq}({\bf r},{\bf r}')\mathcal{E}[{\bf
r}',{\bf r}'';n^{eq}]=\delta(\textbf{r}-\textbf{r}'').
\end{equation}
One can immediately see that this equation is equivalent to the
well-known Ornstein-Zernike equation

\begin{equation}
h({\bf r},{\bf r}') = c({\bf r},{\bf r}') + \int d^3r'' c({\bf
r},{\bf r}'')n^{eq}({\bf r}'') h({\bf r}'',{\bf r}'). \label{oz}
\end{equation}
where $c({\bf r},{\bf r}')$ and $h({\bf r},{\bf r}')$ are,
respectively, the equilibrium value of $c^{(2)}({\bf r},{\bf r}')$
and $h^{(2)}({\bf r},{\bf r}')$. Sometimes approximate chemical
equations of state may be expressed as a ``closure" relation between
these two properties, as in the so-called hyper-netted chain (HNC)
approximation, which writes \cite{mcquarrie}
\begin{equation}
c({\bf r},{\bf r}') = -\beta u(\textbf{r}, \textbf{r}')
 + h({\bf r},{\bf r}') - \ln [1+ h({\bf r},{\bf
r}')]. \label{hnc2}
\end{equation}
Within this approximation one would have to solve self-consistently
Eqs. (\ref{2}), (\ref{oz}),  and (\ref{hnc2}), together with $c[{\bf
r};n^{eq}]=\int d \textbf{r}' c(\textbf{r}, \textbf{r}')n^{eq}
(\textbf{r}')$, for the properties $n^{eq}({\bf r})$, $ c[{\bf
r};n^{eq}]$, $ c({\bf r},{\bf r}')$ and $ h({\bf r},{\bf r}')$,
given the pair potential  $ u({\bf r},{\bf r}')$ and the potential
$\Psi({\bf r})$ of the external field.

\subsection{Non-equilibrium diffusion in colloidal dispersions}\label{VII.2}

In Ref. \cite{eom2} the time-evolution equation for the local
concentration profile $n (\textbf{r},t)$ was derived by
complementing the continuity equation,
\begin{equation}
\frac{\partial n(\textbf{r},t)}{\partial t} = - {\nabla} \cdot
\textbf{j} (\textbf{r},t), \label{continuity}
\end{equation}
with a constitutive relation constructed by defining the particles'
velocity field $\textbf{u} (\textbf{r},t)$ by
\begin{equation}
\textbf{u} (\textbf{r},t)\equiv \textbf{j} (\textbf{r},t)/
n(\textbf{r},t). \label{udrt}
\end{equation}
Thus, $\textbf{u} (\textbf{r},t)$ is the velocity of a particle
representative of the set of particles in a volume $d \textbf{r}$
centered at position $\textbf{r}$. One may then essentially follow
Einstein's proposal \cite{einstein} that the friction force on this
representative particle must be equilibrated, on the average, by the
osmotic force $-\nabla\mu^{in} [{\bf r};n] $ and by the external
force $-\nabla\Psi ({\bf r})$, both included in $-\nabla\mu [{\bf
r};n] $.

The friction force is the sum of the friction due to the supporting
solvent, $-\zeta^0 \textbf{u} (\textbf{r},t)$, and the frictional
effects of the interactions with the other particles,
$-(\Delta\zeta) \textbf{u} (\textbf{r},t)$. The latter, however, may
involve in general spatial and temporal nonlocal effects due to the
collective character of $\textbf{u} (\textbf{r},t)$, so that the
static equilibrium of these forces must actually be written in
general as
\begin{equation}
 \zeta
^0\textbf{u} (\textbf{r},t)+\int_{0}^{t}dt{\acute{}}\int d^3r'\Delta
\zeta (\textbf{r}-\textbf{r}';t-t') \textbf{u} (\textbf{r}',t')
=-\nabla\mu[{\bf r};n(t)]. \label{langevin3}
\end{equation}
whose solution for $\textbf{u} (\textbf{r},t)$ is given by
\begin{equation}\label{udrt2}
\textbf{u} (\textbf{r},t) =-D^0\int_0^t dt' \int d^3r'
b[\textbf{r}-\textbf{r}';t-t']\nabla\beta\mu[{\bf r}';n(t')],
\end{equation}
with the spatially  and temporally non-local collective mobility
kernel $b[\textbf{r}-\textbf{r}';t]$, viewed as the
$(\textbf{r},\textbf{r}')$ element of the ``matrix" $b(t)$, being
the solution of a ``matrix" equation which in Laplace space reads
\begin{equation}\label{bdz}
{b}(z)\circ \left[I+\Delta{\zeta}(z)/\zeta^0\right]=I
\end{equation}
where the matrix product $``\circ"$ means spatial convolution, and
$``I"$ is Dirac's delta function $\delta(\textbf{r}-\textbf{r}')$.
In this equation, the matrix ${b}(z)\equiv \int_0^\infty dt
e^{-zt}{b}(t)$ is the Laplace transform of the matrix $b(t)$, and
similarly for $\Delta\zeta(z)$. The free diffusion coefficient $D^0$
is defined as $D^0\equiv k_BT/\zeta^0$.

Using Eq. (\ref{udrt2}) in $\textbf{j} (\textbf{r},t)= \textbf{u}
(\textbf{r},t)n (\textbf{r},t)$, the continuity equation
(\ref{continuity}) finally leads us to the most general diffusion
equation, namely,
\begin{equation}
\frac{\partial n(\textbf{r},t)}{\partial t} = D^0{\nabla} \cdot
n(\textbf{r},t)\int_0^t dt'\int d^3r' \
b[\textbf{r}-\textbf{r}';t-t']\nabla'\beta\mu[{\bf r}';n(t')].
\label{relaxation3}
\end{equation}
Let us now discuss the use of this equation to describe the
relaxation of the macroscopically observed mean value
$\overline{n}(\textbf{r},t)$ and of the covariance
$\sigma(\textbf{r},\textbf{r}';t)$, as well as of the time-dependent
correlation function $C(t,t')$ of the fluctuations $\delta
n(\textbf{r},t) = n(\textbf{r},t)- \overline{n}(\textbf{r},t)$.

\subsection{Irreversible relaxation of $\overline{n}(\textbf{r},t)$,
$\sigma(\textbf{r},\textbf{r}';t)$, and $C(t,t')$}\label{VII.3}

Following the general format of the relaxation equations presented
in section \ref{VI} (i.e., Eqs. (\ref{releq0repetpp}),
(\ref{sigmadtirrev2}), and (\ref{fluctuations2}), and
(\ref{fluctuations3})), we assume that the spatial and temporal
arguments of the mean local concentration
$\overline{n}(\textbf{r},t)$ describe spatial and temporal
variations of macroscopic scale, so that, for example, in a
quenching process, the variable $t$ is the ageing or waiting time.
In contrast, the thermal fluctuations $\delta n(\textbf{r},t+\tau) =
n(\textbf{r},t+\tau)- \overline{n}(\textbf{r},t)$ vary within
microscopic times denoted by $\tau$ which may be much shorter than
$t$. In a similar manner, we also assume that the spatial variation
of $\overline{n}(\textbf{r},t)$, described by the spatial argument
$\textbf{r}$, occur in much larger spatial scales than the
microscopic spatial variations of the thermal fluctuations $\delta
n(\textbf{r}+\textbf{x},t+\tau)$ indicated in the neighborhood of
$\textbf{r}$ by the spatial coordinate \textbf{x}.  Thus, our
central assumption is that the mean value
$\overline{n}(\textbf{r},t)$ remains approximately uniform and
stationary while the fluctuations vary microscopically within the
finer space and time scales indicated by \textbf{x} and $\tau$. We
express this assumption by describing the macroscopic relaxation of
$\overline{n}(\textbf{r},t)$ by the temporally and spatially
\emph{local} version of Eq. (\ref{relaxation3}). This corresponds to
approximating the generalized mobility kernel
$b[\textbf{r}-\textbf{r}';t-t']$ of this equation by
\begin{equation}
b[\textbf{r}-\textbf{r}';t-t']=b^*(\textbf{r},t)\delta(\textbf{r}-\textbf{r}')2
\delta(t-t'), \label{label1}
\end{equation}
where
\begin{equation}
b^*(\textbf{r},t)\equiv \int d\textbf{x}\int_0^\infty d\tau \
b[\textbf{x},\tau;\textbf{r},t] \label{bast}
\end{equation}
with $b[\textbf{x},\tau;\textbf{r},t]\equiv
b[(\textbf{r}+\textbf{x})-\textbf{r};(t+\tau)-t]$. In this manner we
can write the analog of  Eq. (\ref{releq0repetpp}), i.e., the
diffusion equation for the mean value $\overline{n}(\textbf{r},t)$,
which reads
\begin{equation} \frac{\partial \overline{n}(\textbf{r},t)}{\partial
t} = D^0{\nabla} \cdot \overline{n}(\textbf{r},t) \
b^*(\textbf{r},t)\nabla \left(\beta\mu[{\bf
r};\overline{n}(t)]-\beta\mu^{eq}\right). \label{difeqdl}
\end{equation}

We may now linearize this equation around the equilibrium profile
$n^{eq}({\bf r})$, to get the analog of Eq. (\ref{linearreleqh}) and
(\ref{relaxmatrixh}). From the resulting linearized equation we can
identify the ``matrix" $\mathcal{L}^*
[\textbf{r},\textbf{r}';{\overline n}(t)]$ of Onsager kinetic
coefficients as

\begin{equation}
-\mathcal{L}^* [\textbf{r},\textbf{r}';{\overline  n }(t)] =
D^0{\nabla} \cdot \overline{n}(\textbf{r},t)  \
b^*(\textbf{r},t)\nabla \delta(\textbf{r}-\textbf{r}').
\label{matrixl3}
\end{equation}
and, from Eq. (\ref{sigmadtirrev2}), we can write the relaxation
equation for $\sigma(\textbf{r},\textbf{r}';t)$ as

\begin{eqnarray}
\begin{split}
\frac{\partial \sigma(\textbf{r},\textbf{r}';t)}{\partial t} = &
D^0{\nabla} \cdot \overline{n}(\textbf{r},t) \
b^*(\textbf{r},t)\nabla \int d \textbf{r}_2
\mathcal{E}[\textbf{r},\textbf{r}_2;\overline{n}(t)]
\sigma(\textbf{r}_2,\textbf{r}';t) \\ & +  D^0{\nabla}' \cdot
\overline{n}(\textbf{r}',t) \ b^*(\textbf{r}',t)\nabla' \int d
\textbf{r}_2 \mathcal{E}[\textbf{r}',\textbf{r}_2;\overline{n}(t)]
\sigma(\textbf{r}_2,\textbf{r};t) \\ & -2D^0{\nabla} \cdot
\overline{n}(\textbf{r},t)  \ b^*(\textbf{r},t)\nabla
\delta(\textbf{r}-\textbf{r}'). \label{relsigmadif2}
\end{split}
\end{eqnarray}

Also according to the generalized Onsager scheme, the dynamics of
the fluctuations $\delta n(\textbf{r},t+\tau) \equiv
n(\textbf{r},t+\tau)- \overline{n}(\textbf{r},t)$ are now described
by a stochastic equation with the structure of Eq.
(\ref{fluctuations2}). In our case, this equation is meant to
describe the relaxation of the fluctuations $\delta
n(\textbf{r},t+\tau)$ in the temporal scale described by the time
$\tau$, around the mean value $\overline{n}(\textbf{r},t)$ of the
local concentration at position \textbf{r} and time $t$. The
assumption of local stationarity means that in the time-scale of
$\tau$, $\overline{n}(\textbf{r},t)$ is to be treated as a constant.
Although not explicitly contemplated in the format of Eq.
(\ref{fluctuations2}), but as already indicated above Eq.
(\ref{label1}), here we also add the spatial counterpart of this
simplifying assumption. Thus, we write the fluctuations as $\delta
n(\textbf{r}+\textbf{x},t+\tau) \equiv
n(\textbf{r}+\textbf{x},t+\tau)- \overline{n}(\textbf{r},t)$, where
the argument \textbf{r} of $\overline{n}(\textbf{r},t)$ refer to the
macroscopic resolution of the measured variations of the local
equilibrium profile, whereas the position vector \textbf{x} adds the
possibility of microscopic resolution in the description of the
thermal fluctuations. Defining the fluctuations as the deviations of
the microscopic local concentration profile
$n(\textbf{r}+\textbf{x},t+\tau)$ from the mean value
$\overline{n}(\textbf{r},t)$ indicates that, within the microscopic
spatial variations described by the position vector \textbf{x},
$\overline{n}(\textbf{r},t)$ must be treated as a constant.

With this understanding, we can now proceed to identify the elements
of Eq. (\ref{fluctuations2}) corresponding to our problem. In the
present case, the corresponding antisymmetric matrix
$\omega[\overline{\textbf{a}}(t)]$ vanishes due to time-reversal
symmetry arguments \cite{delrio}. We can then write the matrix
$L[\tau;\overline{\textbf{a}}(t)]$ as the non-markovian and
spatially non-local Onsager matrix implied by the general diffusion
equation in Eq. (\ref{relaxation3}), which must reflect, in
addition, that within the temporal and spatial resolution of the
variables \textbf{x} and $\tau$, the local concentration profile
$\overline{n}(\textbf{r},t)$ remains uniform and stationary. These
assumptions can be summarized by the following stochastic equation
for $\delta n(\textbf{r}+\textbf{x},t+\tau)$

\begin{eqnarray}
\begin{split}\label{fluct}
\frac{\partial \delta n(\textbf{r}+\textbf{x},t+\tau)}{\partial
\tau} =   & D^0\overline{n}(\textbf{r},t){\nabla}_\textbf{x} \cdot
\int_0^\tau d\tau'\int d\textbf{x}_1
b[\textbf{x}-\textbf{x}_1,\tau-\tau';\textbf{r},t]\nabla_{\textbf{x}_1}  \\
& \int d\textbf{x}_2
 \sigma^{-1}(\textbf{x}_1,\textbf{x}_2;t)\delta
n(\textbf{r}+\textbf{x}_2,t+\tau')  +
\textbf{f}(\textbf{r}+\textbf{x},t+\tau),
\end{split}
\end{eqnarray}
where the function $ \sigma^{-1}(\textbf{x},\textbf{x}';t)$ is the
inverse of the covariance $ \sigma(\textbf{x},\textbf{x}';t)$ in the
sense that
\begin{equation}
\int d\textbf{x}'' \ \
\sigma^{-1}(\textbf{x},\textbf{x}'';t)\sigma(\textbf{x}'',\textbf{x}';t)=
\delta(\textbf{x}-\textbf{x}'). \label{sxsinveqi}
\end{equation}
The random term $\textbf{f}(\textbf{r}+\textbf{x},t+\tau)$ of eq.
(\ref{fluct}) is assumed to have zero mean and time correlation
function given by
$<\textbf{f}(\textbf{r}+\textbf{x},t+\tau)\textbf{f}^{\dagger}
(\textbf{r}+\textbf{x}',t+\tau')>=L[\textbf{x}-\textbf{x}',\tau-\tau';\textbf{r},t]$,
with

\begin{equation}\label{nonmarkovmatrixl}
L[\textbf{x}-\textbf{x}',\tau;\textbf{r},t] \equiv
D^0\overline{n}(\textbf{r},t){\nabla}_\textbf{x} \cdot \int
d\textbf{x}_1 b[\textbf{x}-\textbf{x}_1,\tau;
\textbf{r},t]\nabla_{\textbf{x}_1} \delta(\textbf{x}_1-\textbf{x}').
\end{equation}

Similarly, the analog  of Eq. (\ref{fluctuations3}) for the time
correlation function $C_t(\tau)$ is the relaxation equation for
$C(\textbf{x}-\textbf{x}',\tau-\tau';\textbf{r},t) \equiv <\delta
n(\textbf{r}+\textbf{x},t+\tau)\delta
n(\textbf{r}+\textbf{x}',t+\tau')>$, namely,
\begin{eqnarray}
\begin{split}
\frac{\partial C(\textbf{x},\tau;\textbf{r},t) }{\partial \tau} = &
D^0\overline{n}(\textbf{r},t){\nabla}_\textbf{x} \cdot \int_0^\tau
d\tau'\int d\textbf{x}_1 b[\textbf{x}-\textbf{x}_1,\tau-\tau';
\textbf{r},t]\nabla_{\textbf{x}_1} \\ & \int d\textbf{x}_2
\sigma^{-1}(\textbf{x}_1,\textbf{x}_2;t)C(\textbf{x}_2,
\tau';\textbf{r},t). \label{ctdtau2}
\end{split}
\end{eqnarray}

In this manner the generalized theory of non-equilibrium diffusion
just presented writes the relaxation of the mean value
$\overline{n}(\textbf{r},t)$, of the covariance
$\sigma(\textbf{r},\textbf{r}';t)$, and of the time-correlation
function $C(\textbf{x},\tau; \textbf{r},t)$, through Eqs.
(\ref{difeqdl}), (\ref{relsigmadif2}), and (\ref{ctdtau2}),  in
terms of the generalized mobility $b[\textbf{x},\tau; \textbf{r},t]$
or, according to Eq. (\ref{bdz}), in terms of the temporally and
spatially nonlocal friction function $\Delta \zeta [\textbf{x},\tau;
\textbf{r},t]$. These equations constitute the general framework in
which approximations may be introduced to construct a closed system
of equations for the properties involved. This was in fact the main
aim of Ref. \cite{eom2}, and hence, at this point we refer the
reader to that reference for further details.

\section{Concluding remarks}\label{VIII}

In this paper we have proposed a generalization of Onsager's theory
of the time-dependent thermal fluctuations $\delta \textbf{a}(t) =
\textbf{a}(t)- \textbf{a}^{eq}$ around the equilibrium state
$\textbf{a}^{eq}$, to the description of the thermal fluctuations
$\delta \textbf{a}(t) = \textbf{a}(t)- \overline{\textbf{a}}(t)$
around the time-dependent mean value $\overline{\textbf{a}}(t)$ that
relaxes irreversibly towards its most stable equilibrium state
$\textbf{a}^{eq}$ as the solution of a (generally nonlinear)
relaxation equation. The essential results of this generalized
theory were summarized in section \ref{VI}, and consist of the
relaxation equations for the covariance matrix $\sigma (t) =
\overline{\delta \textbf{a}(t)\delta \textbf{a}(t)}$ in Eq.
(\ref{sigmadtirrev2}) and for the two-time correlation function
$C_t(\tau)= \overline{\delta \textbf{a}(t+\tau)\delta
\textbf{a}(t)}$ in Eq. (\ref{fluctuations3}). The time $t$
represents the macroscopic relaxation time that describes the
time-evolution of both, the mean $\overline{\textbf{a}}(t)$ and the
covariance $\sigma (t)$, whereas the time $\tau$ in $C_t(\tau)$
represents the (``microscopic") correlation time of the
fluctuations, as observed at the macroscopic time $t$ after the
system was prepared at the given initial conditions
$\overline{\textbf{a}}(t=0)= {\textbf{a}}^0$ and $\sigma
(t=0)=\sigma^0$.

Eqs. (\ref{releq0repetpp}), (\ref{sigmadtirrev2}), and
(\ref{fluctuations3}), together with the relationship in Eq.
(\ref{okcssppp}), would constitute a closed system of equations if
two fundamental pieces of information were available. The first is
the fundamental thermodynamic relation $S=S[\textbf{a}]$ and the
second refers to the conservative and dissipative kinetic matrices,
$\omega[{\overline{\textbf a}}^0(t)]$ and $L[\tau;{\overline
{\textbf a }}(t),\sigma(t)]$, entering in Eq. (\ref{fluctuations3}).
These two fundamental pieces of information must be provided
externally, and must reflect the specific context of a  particular
relaxation phenomenon. In order to illustrate the use of this
generalized canonical theory by means of a particular application,
in the previous section we described its application to the specific
context of the dynamics of colloidal dispersions. There we first
discussed the main features of the fundamental thermodynamic
relation of these systems and then proposed physical arguments
leading to a generalized diffusion equation.

These are the two elements that had to be provided externally to the
canonical theory. With these two inputs discussed, and following the
script of the canonical theory, we could write the time evolution
equations for the mean local concentration profile ${\overline
n}(\textbf{r},t)$ and for the covariance
$\sigma(\textbf{r},\textbf{r}';t)$ of the thermal fluctuations
$\delta n(\textbf{r},t) \equiv n(\textbf{r},t)- {\overline
n}(\textbf{r},t)$ (Eqs. (\ref{difeqdl}) and (\ref{relsigmadif2}),
respectively). The time evolution of these two properties, as they
relax irreversibly from some arbitrary initial values ${\overline
n}^0(\textbf{r})$ and $\sigma^0(\textbf{r},\textbf{r}')$ to their
equilibrium values ${\overline n}^{eq}(\textbf{r})$ and
$\sigma^{eq}(\textbf{r},\textbf{r}')$, constitute the fundamental
results for the description of the irreversible relaxation of the
system.

We then derived the relaxation equation in Eq. (\ref {ctdtau2}) for
the time-dependent correlation function
$C(\textbf{x},\tau;\textbf{r},t) \equiv \overline{\delta
n(\textbf{r}+\textbf{x},t+\tau)\delta n(\textbf{r},t)}$ in terms of
the generalized time-dependent friction function $\Delta \zeta
[\textbf{x},\tau; \textbf{r},t]$. This is in fact the memory
function of a generalized mobility function $b^*[\textbf{x},\tau;
\textbf{r},t]$ appearing in Eq. (\ref {ctdtau2}) and, in its
Markovian limit, also in Eqs. (\ref{difeqdl}) and
(\ref{relsigmadif2}). Thus, in summary, in the previous section we
expressed the main properties that describe the dynamics of the
non-equilibrium relaxation of a colloidal dispersion, namely,
${\overline n}(\textbf{r},t)$, $\sigma(\textbf{r},\textbf{r}';t)$,
and $C(\textbf{x},\tau;\textbf{r},t)$, in terms of the memory
function $\Delta \zeta [\textbf{x},\tau; \textbf{r},t]$. The
corresponding system of equations constitute the non-equilibrium
generalization of the fundamental equations upon which
non-equilibrium generalization of the SCGLE theory of the dynamics
of equilibrium dispersions can be constructed. In fact, Eq. (\ref
{ctdtau2}) for $C(\textbf{x},\tau;\textbf{r},t)$, together with Eq.
(\ref{bdz}), is the non-equilibrium analog of the memory equation
for the intermediate scattering function $F(k,t)$ in terms of the
so-called irreducible memory function  $\Delta \zeta (k,t)$, as
indicated in detail in Ref. \cite{eom2}. There a proposal is
presented of the full non-equilibrium generalization of the SCGLE
theory of colloid dynamics and of its application to dynamic arrest
phenomena.

In the same reference it is pointed out that the same
non-equilibrium scheme summarized in the previous section contains
as particular cases some results and equations that are important in
various specific contexts. For example, the diffusion equation for
${\overline n}(\textbf{r},t)$ in Eq. (\ref{difeqdl}), in the limit
in which we ignore the memory effects contained in $\Delta \zeta
(k,t)$, happens to coincide with the central general equation of the
recently-developed dynamic density functional theory
\cite{tarazona1, archer}, which has been applied to a variety of
systems, including the description of the irreversible sedimentation
of real and simulated colloidal suspensions
\cite{royalvanblaaderen}. It also coincides in certain circumstances
with an equation for the irreversible relaxation of
$\overline{n}(\textbf{r},t)$ derived by Tokuyama \cite{tokuyama1,
tokuyama2}. On the other hand, the relaxation equation  for the
covariance $\sigma(\textbf{r},\textbf{r}';t)$ in Eq.
(\ref{relsigmadif2}), also in the limit $\Delta \zeta (k,t)=0$, may
also be shown to contain, as a particular case, the fundamental
equation employed in the classical description of the early stages
of spinodal decomposition \cite{cook,langer,dhont,goryachev}.

Thus, the general results of the previous section can be used as the
basis for the extension of the aforementioned theories to the case
in which the memory effects due to the direct interactions,
contained in $\Delta \zeta (k,t)$, cannot be ignored. These effects
are responsible, for example, for the dynamic slowing down of the
system and for its eventual transition to dynamically arrested
conditions. Thus, it is this more general scheme that is expected to
generate the most original predictions including, for example, the
dependence of the glass transition scenario on the cooling rate or
the description of the ageing of the static structure factor and of
the intermediate scattering function after a quenching process.
Specific and more concrete advances in this direction, however, will
be reported separately \cite{pedro3}.

\bigskip

ACKNOWLEDGMENTS: The author acknowledges Pedro Ram\'irez-Gonz\'alez
for stimulating discussions and Rigoberto Ju\'arez-Maldonado,
Alejandro Vizcarra-Rend\'on and Luis Enrique S\'anchez-D\'iaz for
their continued interest in this subject. This work was supported by
the Consejo Nacional de Ciencia y Tecnolog\'{\i}a (CONACYT,
M\'{e}xico), through grant No. 84076.

\end{document}